\RequirePackage{fix-cm}
\documentclass[twocolumn,epjc3]{svjour3}  
\smartqed  
\RequirePackage{graphicx}
\usepackage{times}
\journalname{Eur. Phys. J. A}

\usepackage{amsmath,amssymb}
\usepackage{bm}
\usepackage{braket}
\begin{document}

\title{Compositeness of hadrons, nuclei, and atomic systems
}


\author{Tomona Kinugawa\thanksref{addr1,addr2,e1}
        \and
        Tetsuo Hyodo\thanksref{addr1,e2} 
}

\thankstext{e1}{e-mail: kinugawa-tomona@ed.tmu.ac.jp}
\thankstext{e2}{e-mail: hyodo@tmu.ac.jp}


\institute{Department of Physics, Tokyo Metropolitan University, Hachioji 192-0397, Japan \label{addr1}
           \and
           \emph{Present Address: Nishina Center for Accelerator-Based Science, RIKEN, Wako 351-0198, Japan \label{addr2}}
}

\date{Received: 30 September 2024 / Accepted: 25 March 2025}

\maketitle

\begin{abstract}
Recent observations of exotic hadrons have stimulated the theoretical investigation of the internal structure of hadrons. While all hadrons are ultimately composed of quarks and gluons bound by the strong interaction, quark clustering phenomena can generate hadronic molecules---weakly bound systems of hadrons---which are expected to emerge near two-hadron thresholds. However, it should be noted that a pure hadronic molecule is not realized, as the strong interaction induces mixing with other possible configurations. The compositeness of hadrons has been developed as a promising concept to quantitatively characterize the fraction of the hadronic molecular component. Here we summarize the modern understanding of the compositeness to study the internal structure of hadrons and review the application of the compositeness to various quantum systems in different energy scales, such as nuclei and atomic systems, in addition to hadrons. 

\keywords{Compositeness \and structure of hadrons \and clustering phenomena \and low-energy universality \and effective field theory}
\end{abstract}
%

\section{Introduction}\label{sec:intro}

In recent hadron spectroscopy, thanks to the rapid experimental developments, the number of observed hadrons continues to increase~\cite{ParticleDataGroup:2024cfk}. In particular, it is remarkable that many candidates of exotic hadrons have been found, starting from the observation of the charmonium-like state $X(3872)$~\cite{Belle:2003nnu} to the recent discovery of the pentaquark $P_{c}$~\cite{LHCb:2015yax,LHCb:2019kea} and the tetraquark $T_{cc}$~\cite{LHCb:2021vvq,LHCb:2021auc}. These newly observed hadrons do not fit well in the traditional classification of hadrons, namely, $\bar{q}q$ mesons and $qqq$ baryons~\cite{Hosaka:2016pey,Guo:2017jvc,Brambilla:2019esw,Chen:2022asf}. This has triggered an intensive discussion on the internal structure of hadrons such as multiquark states, hadronic molecules, and gluon hybrids.

Among others, hadronic molecules are of particular interest from the perspective of clustering phenomena. In contrast to ordinary hadrons, which are made from quarks and gluons by quantum
chromodynamics (QCD) interactions, the main building blocks of hadronic molecules are the ground-state hadrons (constituent hadrons), and the hadron-hadron interaction is the driving force of self-bound systems. Because these constituent hadrons are composed of quarks, hadronic molecules can be regarded as quark clusters. As a universal feature of clustering~\cite{PTPS52.89}, it is expected that hadronic molecules are formed if the binding energy of the two hadrons is much smaller than the binding of quarks in the constituent hadrons. The property of weak binding allows the constituent hadrons to keep their identity as hadrons. But the clustering of quarks has two peculiar features which are not shared by the clustering of other quantum systems in different hierarchies such as nuclei and atomic systems:
\begin{itemize}
\item Color confinement: quarks are confined in hadrons, so the ``binding energy'' of quarks in a hadron is not well defined. Also, the clustering into the non-color-singlet subunit is (at least asymptotically) not possible. 
\item Pair creation and annihilation: quark-antiquark pairs can be created and annihilated by the strong interaction. This leads to the mixing of states with different numbers of particles, such as $\bar{q}q\leftrightarrow \bar{q}\bar{q}qq$.
\end{itemize}
These features make the discussion of hadronic molecules interesting but also challenging. 

It should be emphasized that hadronic molecules cannot be distinguished from normal hadrons by the conserved quantum numbers. For a discussion of hadronic molecules, therefore, we need to introduce some measure to quantify the internal structure. In this context, the concept of compositeness has recently drawn significant attention. The notion of compositeness appeared in discussions in the 1960s to distinguish between composite and elementary particles~\cite{Weinberg:1962hj,Weinberg:1963zza,Weinberg:1964zza,Weinberg:1965zz}. In the recent revival of the concept of compositeness for exotic hadrons, it is important to incorporate the unstable nature of hadrons, because most of the candidates for hadronic molecules are unstable against  strong decay. 

In this paper, we review the theoretical foundation of the study of the compositeness of hadrons, including recent developments. A brief survey from 2013 can be found in a review article~\cite{Hyodo:2013nka} written by one of the present authors (see also Refs.~\cite{Guo:2017jvc,Oller:2017alp,vanKolck:2022lqz} from other viewpoints). Since then, the notion of the compositeness has been widely applied in hadron physics. Moreover, thanks to low-energy universality, the compositeness can also be used for systems in different hierarchies.

This paper is organized as follows. In Sect.~\ref{sec:history}, we summarize the history of the study of the compositeness of hadrons. The basic formulation of the compositeness is introduced in Sect.~\ref{sec:formulation} in the framework of non-relativistic quantum mechanics. We emphasize the model dependence of the compositeness in various expressions and present the generalization to higher partial waves and to the coupled-channel scattering. Section~\ref{sec:WBR} deals with the weak-binding limit where the structure of a weakly bound state in an $s$ wave is determined in a model-independent manner. We demonstrate how model dependence of the compositeness is suppressed in this limit. The results in Sects.~\ref{sec:formulation} and \ref{sec:WBR} are revisited from the viewpoint of the effective field theory in Sects.~\ref{sec:EFT}. We demonstrate that some issues in sections~\ref{sec:formulation} and \ref{sec:WBR} can be well understood in the formulation with field theory. The generalization to unstable resonances is discussed in Sects.~\ref{sec:unstable}, and several interpretation schemes of the complex compositeness are summarized. Applications to various physical systems, including hadrons, nuclei, and atomic systems, are presented in Sects.~\ref{sec:applications}.

\section{Brief history}\label{sec:history}

One of the central problems in particle physics in the 1960s was in distinguishing the composite particles from elementary ones. A series of works by Weinberg~\cite{Weinberg:1962hj,Weinberg:1963zza,Weinberg:1964zza,Weinberg:1965zz} appeared in this context, and it was shown that the vanishing of the field renormalization constant ($Z=0$) could be used as a criterion to characterize the composite particle, and this constant could be related to the scattering length and the effective range for a weakly bound state (weak-binding relation), with the successful application to the deuteron. As mentioned in  Weinberg's paper, the $Z=0$ condition for the composite particle was noted in various approaches~\cite{Jouvet:1957ja,NC18.466,Vaughn:1961poz,NC24.870,Salam:1962ap,Salam:1963sa,NC25.1135,PTP29.877,PR136.B816}. The non-relativistic limit, as in the Weinberg's paper~\cite{Weinberg:1965zz}, was discussed in Ref.~\cite{NC25.1135} and the application to the deuteron using the low-energy scattering parameters was attempted in Ref.~\cite{PTP29.877}. Nevertheless, Weinberg's work~\cite{Weinberg:1965zz} provides an important step forward to establishing the weak-binding relation as a model-independent relation of the observables and the internal structure of the bound state.

The discussion of the compositeness in the 1960s was summarized in a review article~\cite{Hayashi:1967bjx}. Since then, the compositeness condition $Z=0$ has been mainly applied in high-energy physics~\cite{Tirapegui:1971tsc,Lemmon:1975yi,Shizuya:1979bv} rather than hadron physics. An interesting application to hadron physics using the $K$-matrix approach can be found in Ref.~\cite{Rajasekaran:1972gs}, where various hadrons, including $\Lambda(1405)$, are studied. Also, the pole counting method in Refs.~\cite{Morgan:1990ct,Morgan:1992ge,Morgan:1993td} is closely related to Weinberg's discussion.

As mentioned in Sect.~\ref{sec:intro}, in this century, the internal structure of hadrons has drawn attention because of the observation of various exotic hadron candidates. In the application of the compositeness to exotic hadrons, one is immediately faced with the problem of the unstable nature of these hadrons, where subtleties arise in the definition and the interpretation of the compositeness (see Sect.~\ref{sec:unstable}). Various approaches have been proposed to deal with the unstable resonances. Recent interest in the compositeness and Weinberg's work was triggered by the study of $f_{0}(980)$ and $a_{0}(980)$ using the integration of the spectral function~\cite{Baru:2003qq}. The framework in Ref.~\cite{Baru:2003qq} was further developed in Refs.~\cite{Kalashnikova:2009gt,Baru:2010ww,Hanhart:2011jz,Wang:2022vga} in relation to the Castillejo--Dalitz--Dyson (CDD) zeros~\cite{Castillejo:1956ed,PR124.264}, Flatt\'e amplitude, and the coupled-channel effects. The expression of the compositeness using the residue of the resonance pole was discussed in Refs.~\cite{Gamermann:2009uq,Hyodo:2011qc,Aceti:2012dd,Aceti:2014ala}, which turns out to be useful for practical applications~\cite{Sekihara:2014kya,Nagahiro:2014mba}. The residue of the pole can be interpreted by the rank-1 projection operator to the state expressed by the pole~\cite{Guo:2015daa} and by the number operator~\cite{Oller:2017alp}. Generalization of the compositeness of bound states to near-threshold resonances using the effective range expansion was introduced in Ref.~\cite{Hyodo:2013iga}. The compositeness with the effective range expansion is further discussed in Ref.~\cite{Matuschek:2020gqe} for virtual states with a different criterion and in Ref.~\cite{Baru:2021ldu} including various modifications in a realistic situation. The compositeness of the bound state in the limit of the vanishing of the binding energy is discussed in relation with the near-threshold mass scaling law~\cite{Hyodo:2014bda,Hanhart:2014ssa}. The structure of the state with a small but finite binding energy is further studied in Refs.~\cite{Sazdjian:2022kaf,Lebed:2022vks,Hanhart:2022qxq,Kinugawa:2023fbf} in relation to the fine-tuning of parameters. Generalization of the weak-binding relation to unstable states was performed by using effective field theories (EFTs)~\cite{Kamiya:2015aea,Kamiya:2016oao}. Further discussion on the EFT approach can be found in Ref.~\cite{vanKolck:2022lqz}. Recently, several methods of finite-range corrections were
discussed in relation to the quantitative reanalysis of the compositeness of the deuteron~\cite{Li:2021cue,Baru:2021ldu,Song:2022yvz,Albaladejo:2022sux,Kinugawa:2022fzn,Yin:2023wls}.

\section{Formulation of compositeness}\label{sec:formulation}

In this section, we discuss the compositeness of a bound state in non-relativistic quantum mechanics, basically following the discussion in the original paper~\cite{Weinberg:1965zz} and the review article~\cite{Hyodo:2013nka}. We consider the short-range two-body interaction which is spherically symmetric, so that the scattering occurs independently in each partial wave. 

\subsection{Hamiltonian and eigenstates}\label{subsec:Hamiltonian}

For simplicity, we first focus on the single-channel $s$-wave two-body scattering with one bound state, without any internal degrees of freedom such as spin and flavor. We consider the non-relativistic quantum system governed by the Hamiltonian $H$. The eigenstates of $H$ satisfy the following Schr\"odinger equation:
\begin{align}
   H\ket{\bm{p},\pm} 
   &= E_{p}\ket{\bm{p},\pm} ,\quad
   H\ket{B}
   = -B\ket{B} ,
   \label{eq:Schroedingereq} \\
   E_{p}&=\frac{\bm{p}^{2}}{2\mu} ,
\end{align}
where $\ket{\bm{p},\pm}$ are the asymptotic scattering solutions with continuous eigenvalues, $\mu$ is the reduced mass of the scattering channel, and $\ket{B}$ is the bound state solution with binding energy $B$. We consider the bound state having the same quantum numbers as the $s$-wave scattering states. The normalization of the states is given by
\begin{align}
   \braket{\bm{p}^{\prime},\pm|\bm{p},\pm} 
   &=\delta(\bm{p}^{\prime}-\bm{p}), \quad
   \braket{B|B}=1 ,
   \label{eq:normalizationfull} \\
   \braket{\bm{p},\pm | B}
   &=0 .
\end{align}
Asymptotic completeness guarantees the relation~\cite{Taylor}
\begin{align}
   1
   &=\int d\bm{p} \ket{\bm{p},\pm}\bra{\bm{p},\pm}+\ket{B}\bra{B}
   \label{eq:completenessfull} .
\end{align}

The starting point for the discussion of the compositeness is to separate the Hamiltonian $H$ into the free part $H_{0}$ and the interaction $V$:
\begin{align}
   H&=H_{0}+V . \label{eq:H0pV}
\end{align}
We assume that the eigenstates of the free Hamiltonian $H_{0}$ consist of the continuum of the scattering states $\ket{\bm{p}}$ and a discrete eigenstate $\ket{B_{0}}$ with
\begin{align}
   H_{0}\ket{\bm{p}}=E_{p}\ket{\bm{p}} ,
   \quad
   H_{0}\ket{B_{0}}=\omega_{0}\ket{B_{0}} ,
   \label{eq:SchroedingereqH0}
\end{align}
where $\omega_{0}$ is the bare energy of the discrete state $\ket{B_{0}}$. The orthonormal and completeness relations are given by
\begin{align}
   \braket{\bm{p}^{\prime}|\bm{p}} 
   &=\delta(\bm{p}^{\prime}-\bm{p}), \quad
   \braket{B_{0}|B_{0}}=1, \quad
   \braket{\bm{p} | B_{0}}=0
   \label{eq:normalizationH0}, \\
   1
   &=\int d\bm{p} \ket{\bm{p}}\bra{\bm{p}}+\ket{B_{0}}\bra{B_{0}}
   \label{eq:completenessH0} .
\end{align}
The interaction $V$ provides the transitions between the eigenstate, except for $\braket{B_{0}|V|B_{0}}=0$.

The decomposition~\eqref{eq:H0pV} and the completeness relation~\eqref{eq:completenessH0} require careful explanations. First, we note that the decomposition of $H$ into $H_{0}$ and $V$ is not unique. For instance, it is possible to construct $H_{0}$ and $V$ which satisfy \eqref{eq:completenessH0} without the $\ket{B_{0}}$ term, with physics determined by $H$ being unchanged. It is also possible to construct $H_{0}$ with multiple bare states (see Sect.~\ref{subsec:coupledchannel}). Indeed, the definition of ``free'' $H_{0}$ determines the basis~\eqref{eq:completenessH0} for expanding the physical bound state $\ket{B}$. This non-uniqueness of the basis introduces the model dependence of the compositeness, which we will discuss later in Sect.~\ref{subsec:modeldependence}. One may feel awkward about the origin of $\ket{B_{0}}$ in Eq.~\eqref{eq:completenessH0}. In the context of hadron physics, one may conceptually switch off the hadron-hadron interactions while keeping the confinement dynamics intact, so that a color singlet state having the same quantum numbers as the scattering states $\ket{\bm{p}}$ is formed as $\ket{B_{0}}$. In this sense, $\ket{B_{0}}$ is a remnant of the coupled channels, which are not explicitly considered here. Some concrete examples for the interpretation of $\ket{B_{0}}$ are given in Ref.~\cite{Hyodo:2013nka}. In Sect.~\ref{sec:EFT}, we explicitly construct the state vectors which satisfy Eq.~\eqref{eq:completenessH0} in the framework of effective field theory. In fact, this formulation is essentially equivalent to the time-honored Lee model~\cite{Lee:1954iq,Hyodo:2013nka}. 

\subsection{Definition of compositeness}\label{subsec:definition}

The completeness relation~\eqref{eq:completenessH0} can be used to decompose the physical bound state $\ket{B}$ as a linear combination of the scattering states $\ket{\bm{p}}$ and the bare state $\ket{B_{0}}$:
\begin{align}
   \ket{B}
   =\int d\bm{p}\ \chi(\bm{p})\ket{\bm{p}}+c\ket{B_{0}} , \\
   \chi(\bm{p})=\braket{\bm{p}|B} ,\quad
   c=\braket{B_{0}|B} ,
   \label{eq:cchi}
\end{align}
where $\chi(\bm{p})$ is the momentum representation of the bound state wave function and $c$ is the overlap of the bare state $\ket{B_{0}}$ and the bound state $\ket{B}$. The compositeness $X$ (the elementarity $Z$) is defined as the probability of finding the scattering states $\ket{\bm{p}}$ (the bare state $\ket{B_{0}}$) in the physical bound state $\ket{B}$ as
\begin{align}
   X
   &= \int d\bm{p}\ |\braket{\bm{p}|B}|^{2}
   =
   \int d\bm{p}\ |\chi(\bm{p})|^{2}\geq 0,
   \label{eq:compositeness} \\
   Z
   &=|\braket{B_{0}|B}|^{2}=
   |c|^{2}\geq 0 .
   \label{eq:elementarity} 
\end{align}
Using the normalization condition~\eqref{eq:normalizationfull} and the completeness relation~\eqref{eq:completenessH0}, we find
\begin{align}
   X
   +   Z
   &=1 .
\end{align}
Namely, $X$ and $Z$ are real and nonnegative, and their sum is unity. With these properties, we can interpret $X$ and $Z$ as probabilities. Because $Z$ represents the renormalization of the wave function of the bare state $\ket{B_{0}}$, $Z$ is also called the field (wave function) renormalization constant.

The above formulation can be concisely expressed by the two-component Hamiltonian~\cite{Baru:2003qq,Baru:2010ww,Hyodo:2014bda,Guo:2017jvc}
\begin{align}
   \begin{pmatrix}
   H_{0} & V \\
   V & H_{0} + V
   \end{pmatrix}
   \ket{\Psi}
   &= E\ket{\Psi} ,
   \label{eq:twocomponent}
\end{align}
with
\begin{align}
   \ket{\Psi}
   &=
   \begin{pmatrix}
   c_{E}\ket{B_{0}} \\
   \int d\bm{p}\ \chi_{E}(\bm{p})\ket{\bm{p}} 
   \end{pmatrix}
   , 
\end{align}
where the coefficients $c_{E}$ and $\chi_{E}(\bm{p})$ reduce to $c$ and $\chi(\bm{p})$ in Eq.~\eqref{eq:cchi} for $E=-B$. This Hamiltonian is equivalent to the single resonance approach for the Feshbach resonance~\cite{Kohler:2006zz}. In fact, if we define the operators $Q$ and $P$ as
\begin{align}
   P
   &=
   \int d\bm{p} \ket{\bm{p}}\bra{\bm{p}}, \quad
   Q
   =
   \ket{B_{0}}\bra{B_{0}},
   \label{eq:projections}
\end{align}
these satisfy the properties of the projection operators $P+Q=1, PQ=QP=0, P^{2}=P, Q^{2}=Q$, thanks to Eq.~\eqref{eq:normalizationH0}. In this case, the compositeness $X$ (elementarity $Z$) is interpreted as the expectation value of the projection operator to $P$ space (to $Q$ space)~\cite{Miyahara:2018onh}
\begin{align}
   X
   &=
   \bra{B}P\ket{B},\quad
   Z
   =
   \bra{B}Q\ket{B}.
\end{align}

\subsection{Expressions of compositeness}\label{subsec:expressions}

For the actual calculation of the compositeness, it is useful to translate the definition~\eqref{eq:compositeness} into different expressions. Using the Schr\"odinger equation~\eqref{eq:Schroedingereq}, the compositeness $X$ can be expressed by the energy integral~\cite{Hyodo:2013nka}
\begin{align}
   X
   &=
   \int d\bm{p}\frac{|\braket{\bm{p}|V|B}|^{2}}{(E_{p}+B)^{2}} 
   \nonumber \\
   &=
   4\pi\sqrt{2\mu^{3}}g_{\rm th}^{2}\int_{0}^{\infty} dE\frac{\sqrt{E}|F(E)|^{2}}{(E+B)^{2}} ,
   \label{eq:Xatpole}
\end{align}
where $F(E)$ is the transition form factor of the bound state $\ket{B}$ to the scattering states defined as
\begin{align}
   \braket{\bm{p}|V|B} = g_{\rm th}F(E_{p})
   \label{eq:formfactor} ,
\end{align}
with the normalization $F(0)=1$. Thus, $g_{\rm th}$ is the coupling constant of the bound state and the scattering state at the threshold ($E_{p}=0$). Note that an $s$-wave bound state is assumed here, and the generalization to higher partial waves will be discussed in Sect.~\ref{subsec:higher}. As we will see below, Eq.~\eqref{eq:Xatpole} is the expression of the compositeness $X$ at the bound state pole of the scattering amplitude.

The compositeness can also be expressed by the scattering amplitude. The $s$-wave on-shell $t$-matrix $t(E_{p})$ is defined by the matrix element of the $T$-operator as
\begin{align}
   t(E_{p})
   &=\left.\bra{\bm{p}^{\prime}}T(E_{p}+i0^{+})\ket{\bm{p}}\right|_{E_{p^{\prime}}=E_{p}} .
\end{align}
With the expression of the $T$-operator (Chew-Goldberger solution)
\begin{align}
   T(z)
   &=V+V\frac{1}{z-H}V ,
\end{align}
and the completeness relation~\eqref{eq:completenessfull}, one obtains the expression of the off-shell $t$-matrix
\begin{align}
   &\quad \bra{\bm{p}^{\prime}}T(z)\ket{\bm{p}}\nonumber \\
   &=\bra{\bm{p}^{\prime}}V\ket{\bm{p}}+\frac{
   \bra{\bm{p}^{\prime}}V\ket{B}\bra{B}V\ket{\bm{p}}}{z+B}\nonumber \\
   &\quad 
   +\int d\bm{q}
   \frac{
   \bra{\bm{p}^{\prime}}T(E_{q}+i0^{+})\ket{\bm{q}}\bra{\bm{q}}T(E_{q}-i0^{+})\ket{\bm{p}}
   }{z-E_{q}} .
   \label{eq:offshellscatteringamplitude} 
\end{align}
A solution of this integral equation can be constructed using the on-shell $t$-matrix satisfying~\cite{Weinberg:1965zz,Hyodo:2013nka}
\begin{align}
   t(E)
   &=v+\frac{g_{\rm th}^{2}|F(E)|^{2}}{E+B}
   +4\pi\sqrt{2\mu^{3}}\int_{0}^{\infty}dE^{\prime}
   \frac{\sqrt{E^{\prime}}|t(E^{\prime})|^{2}}{E-E^{\prime}+i0^{+}},
   \label{eq:scatteringamplitude} \\
   v
   &=\bra{\bm{p}}V\ket{\bm{p}} .
   \label{eq:vmatrixelement}
\end{align}
Because the second term in Eq.~\eqref{eq:scatteringamplitude} is the bound state pole contribution to the scattering amplitude, it is now clear that Eq.~\eqref{eq:Xatpole} is the expression at the bound state pole. The residue of the bound state pole is related to the coupling constant squared $g^{2}$ at $E=-B$:
\begin{align}
   g^{2}
   &=
   \lim_{E\to -B}(E+B)t(E)
   =g_{\rm th}^{2}|F(-B)|^{2} .
   \label{eq:residue}
\end{align}
Namely, the residue of the pole is given by the value of the form factor at $E=-B$ and $g_{\rm th}$. Solving Eq.~\eqref{eq:scatteringamplitude} for $g_{\rm th}^{2}|F(E)|^{2}/(E+B)$ and substituting it into Eq.~\eqref{eq:Xatpole}, we obtain the compositeness expressed by the scattering amplitude~\cite{Hyodo:2013nka}
\begin{align}
   X
   &=4\pi\sqrt{2\mu^{3}}
   \int_{0}^{\infty}dE\frac{\sqrt{E}}{E+B}
   \Biggl[
   t(E)-v \nonumber \\
   &\quad 
   -4\pi\sqrt{2\mu^{3}}
   \int_{0}^{\infty}dE^{\prime}
   \frac{\sqrt{E^{\prime}}|t(E^{\prime})|^{2}}{E-E^{\prime}+i0^{+}}
   \Biggr] .
   \label{eq:Xbyt}
\end{align}
Note that the imaginary part of $t(E)$ and that of the $E^{\prime}$ integration cancel each other out so that the compositeness is obtained as a real number~\cite{Hyodo:2013nka}.

\subsection{Model dependence of compositeness}\label{subsec:modeldependence}

At this point, we note that the compositeness $X$ is a model-dependent quantity. In other words, $X$ cannot be directly determined from the experimental observables, except for the special case of the weak-binding limit, which will be discussed in Sect.~\ref{sec:WBR}. It is instructive to summarize the origin of the model dependence of the compositeness in the expressions~\eqref{eq:Xatpole} and \eqref{eq:Xbyt}. 

In general, observable quantities, obtained from the on-shell scattering amplitude, are said to be model-independent (see e.g. Ref.~\cite{Epelbaum:2008ga}). In an ideal situation, the observable quantities can be determined uniquely by experiments. In Eqs.~\eqref{eq:Xatpole} and \eqref{eq:Xbyt}, the binding energy $B$ (pole of the scattering amplitude) and the on-shell $t$-matrix $t(E)$ are the model-independent quantities in this sense.

On the other hand, there are also model-dependent quantities in Eqs.~\eqref{eq:Xatpole} and \eqref{eq:Xbyt}. For instance, the transition form factor $g_{\rm th}F(E_{p})$ in Eq.~\eqref{eq:formfactor} is related to the off-shell property of the bound state $\ket{B}$ (note that in general, $E_{p}\neq -B$), and therefore is not an observable.\footnote{We note that the residue of the on-shell scattering amplitude $g^{2}$ in Eq.~\eqref{eq:residue} can be determined in a model-independent manner. The value of $g_{\rm th}$ and the functional form of $F(E)$ are model-dependent, under the conditions $F(0)=1$ and $g_{\rm th}^{2}|F(-B)|^{2}=g^{2}$.} In other words, it is possible to modify the form factor with the observables being unchanged. In Eq.~\eqref{eq:Xbyt}, the matrix element $v=\bra{\bm{p}}V\ket{\bm{p}}$ depends on the choice of the decomposition $H=H_{0}+V$, as we mentioned in Sect.~\ref{subsec:Hamiltonian}. Depending on the choice of the bare Hamiltonian, the matrix element $v$ can be modified without changing the physics described by $H$. We summarize the classification 
in Table~\ref{tab:modeldep}.

In this way, $g_{\rm th}F(E)$ in Eq.~\eqref{eq:Xatpole} and $v$ in Eq.~\eqref{eq:Xbyt} cannot be uniquely determined by the observables, and they are the source of the model dependence of the compositeness $X$. It should be emphasized that the compositeness $X$ is inherently a model-dependent quantity in a general situation. In Sect.~\ref{sec:WBR}, we discuss how this model dependence disappears in the exceptional case of the weak-binding limit. 

\begin{table}
\caption{Summary of model dependence of physical quantities in Sect.~\ref{subsec:expressions}.}
\label{tab:modeldep}       
\begin{tabular}{ll}
\hline\noalign{\smallskip}
Classification & Quantities   \\
\noalign{\smallskip}\hline\noalign{\smallskip}
Model dependent non-observables & $v$, $F(E)$, $g_{\rm th}$, $X$, $Z$   \\
Model independent observables & $t(E)$, $B$, $g$ \\
\noalign{\smallskip}\hline
\end{tabular}
\end{table}

\subsection{Higher partial waves}\label{subsec:higher}

So far we have considered the case where the bound state couples with the $s$-wave scattering states. The compositeness of the bound state in higher partial waves can also be defined, as discussed in Refs.~\cite{Aceti:2012dd,Aceti:2014ala,Guo:2015daa,Sekihara:2015gvw,Sekihara:2016xnq,Oller:2017alp}. Let us consider the same setup as in Sect.~\ref{subsec:Hamiltonian} but with a bound state in the $\ell$-th partial wave. Because the discussion in Sect.~\ref{subsec:Hamiltonian} and \ref{subsec:definition} is independent of the partial wave of the scattering states, the definition of the compositeness~\eqref{eq:compositeness} and the elementarity~\eqref{eq:elementarity} remain unchanged for arbitrary $\ell$. 

The angular momentum dependence appears in the expressions in Sect.~\ref{subsec:expressions}. Considering the small $|\bm{p}|$ behavior of the form factor, we generalize Eq.~\eqref{eq:formfactor} for an arbitrary $\ell$ with the spherical harmonics $Y_{\ell}^{m}$ as~\cite{Sekihara:2015gvw}
\begin{align}
   \braket{\bm{p}|V|B}
   =g_{\rm th}\sqrt{4\pi}Y_{\ell}^{m}(\hat{p})F(E_{p})|\bm{p}|^{\ell}  ,
   \label{eq:formfactorell}
\end{align}
which reduces to Eq.~\eqref{eq:formfactor} for $\ell=0$.\footnote{Note that the dimension of the bare coupling constant $g_{\rm th}$ depends on $\ell$.} Factorizing $|\bm{p}|^{\ell}$, we can normalize the form factor as $F(0)=1$. In this case, the expression of the compositeness is given by
\begin{align}
   X
   &=
   4\pi\sqrt{2\mu^{3}}g_{\rm th}^{2}
   \int_{0}^{\infty} dE\frac{
   \sqrt{E}(2\mu E)^{\ell}|F(E)|^{2}}{(E+B)^{2}} .
   \label{eq:Xatpoleell}
\end{align}
Namely, the only difference from Eq.~\eqref{eq:Xatpole} is the additional $(2\mu E)^{\ell}$ factor in the integrand. Nevertheless, this factor provides a crucial distinction between $\ell=0$ and $\ell\neq 0$ in the weak-binding limit.

\subsection{Coupled-channel case}\label{subsec:coupledchannel}

In most problems of hadron physics, there are multiple scattering channels that couple with the system of interest. In other words, we need to deal with the coupled-channel problem. Let us consider that there are $I$ scattering channels. The scattering states $\ket{\bm{p},\pm}$ in Eq.~\eqref{eq:Schroedingereq} and $\ket{\bm{p}}$ in Eq.~\eqref{eq:SchroedingereqH0} should be labeled by the index $i=1,2,\dotsb, I$. As we mentioned in Sect.~\ref{subsec:Hamiltonian}, the number of discrete eigenstates in the free Hamiltonian is arbitrary, and therefore, we assume that there are $N$ bare states. In this setup, the eigenstates of the free Hamiltonian now read
\begin{align}
   H_{0}\ket{\bm{p},i}&=E_{p,i}\ket{\bm{p},i}  \quad (i=1,\dotsb,I), \\
   H_{0}\ket{B_{0,n}}&=\omega_{0,n}\ket{B_{0,n}} \quad (n=1,\dotsb,N) , \\
   E_{p,i}
   &=\frac{\bm{p}^{2}}{2\mu_{i}}+E_{{\rm th},i} ,
\end{align}
where $E_{{\rm th},i}$ is the threshold energy of channel $i$, with $E_{{\rm th},1}=0$ and $E_{{\rm th},i+1}\geq E_{{\rm th},i}$.\footnote{Here we consider only the two-body scattering channels labeled by single relative momentum $\bm{p}$. It is also possible to consider the $n$-body scattering states which are labeled by $n-1$ relative momenta with suitable normalizations. The following equations are modified accordingly with $n-1$ momentum integrations, and so on.} The orthonormal conditions for these states are given by
\begin{align}
   \braket{\bm{p}^{\prime},i|\bm{p},j} 
   &=\delta(\bm{p}^{\prime}-\bm{p})\delta_{ij}, \quad
   \braket{B_{0,n}|B_{0,m}}=\delta_{nm}, \nonumber \\
   \braket{\bm{p},i | B_{0,n}}
   &=0
   \label{eq:normalizationH0multi}, 
\end{align}
and the completeness relation is 
\begin{align}
   1
   &=
   \sum_{i=1}^{I}\int d\bm{p} \ket{\bm{p},i}\bra{\bm{p},i}
   +\sum_{n=1}^{N}\ket{B_{0,n}}\bra{B_{0,n}}
   \label{eq:completenessH0multi} .
\end{align}

In the physical spectrum given by the eigenstates of $H$, there can be multiple bound states. In this case, the compositeness is defined for each bound state~\cite{Hyodo:2013nka}. We now focus on one specific bound state $\ket{B}$ in the spectrum of $H$, and expand it using the basis~\eqref{eq:completenessH0multi} as
\begin{align}
   \ket{B}
   =\sum_{i=1}^{I}\int d\bm{p}\ \chi_{i}(\bm{p})\ket{\bm{p}}
   +\sum_{n=1}^{N}c_{n}\ket{B_{n,0}} , \\
   \chi_{i}(\bm{p})=\braket{\bm{p},i|B} ,\quad
   c_{n}=\braket{B_{0,n}|B} .
   \label{eq:cchimulti}
\end{align}
Generalization of Eqs.~\eqref{eq:compositeness} and \eqref{eq:elementarity} is 
\begin{align}
   X_{i}
   &= \int d\bm{p}\ |\braket{\bm{p},i|B}|^{2}
   =
   \int d\bm{p}\ |\chi_{i}(\bm{p})|^{2}\geq 0,
   \label{eq:compositenessmulti} \\
   Z_{n}
   &=|\braket{B_{0,n}|B}|^{2}=
   |c_{n}|^{2}\geq 0.
   \label{eq:elementaritymulti} 
\end{align}
Namely, the compositeness is defined for each scattering channel $i$, and the elementarity is defined for each bare state $n$. It is now clear from Eq.~\eqref{eq:completenessH0multi} that the sum of all $X_{i}$ and $Z_{n}$ is unity:
\begin{align}
   \sum_{i=1}^{I}X_{i}
   +   \sum_{n=1}^{N}Z_{n}
   &=1 
   \label{eq:normalizationmulti} .
\end{align}
In this way, the compositeness $X_{i}$ (the elementarity $Z_{n}$) is interpreted as the probability of finding the scattering state $\ket{\bm{p},i}$ (the bare state $\ket{B_{0,n}}$) in the physical bound state $\ket{B}$. $X_{i}$ and $\sum_{i}X_{i}$ are sometimes called the partial compositeness and the total compositeness, respectively~\cite{Guo:2015daa}, in analogy with the partial and total decay widths.

\section{Weak-binding relations}\label{sec:WBR}

We now show that the compositeness $X$ of an $s$-wave bound state can be determined in a model-independent manner in the weak-binding limit. Again, we follow the discussion in the original paper~\cite{Weinberg:1965zz} and the review article~\cite{Hyodo:2013nka}.

\subsection{Typical scales}

Defining the weak-binding limit requires the knowledge of the energy scale of the system of interest, in addition to the binding energy $B$. For instance, a bound state with $B=1$ MeV may be regarded as a weakly bound state in hadron physics, but it is certainly not in the atomic physics. To quantify the ``weak-binding limit,'' we need to introduce a typical energy scale of the system $E_{\rm typ}$, with which the binding energy $B$ is compared. In the original paper~\cite{Weinberg:1965zz}, it is denoted as $E_{0}$ and is estimated by $\sim m_{\pi}^{2}/(2\mu)$, because the $\pi$ exchange in the nuclear force gives a momentum scale $\sim m_{\pi}$. In this way, the energy scale $E_{\rm typ}$ depends on the microscopic details of the two-body system. In the framework of the effective field theory in Sect.~\ref{sec:EFT}, $E_{\rm typ}$ is related to the ultraviolet cutoff in the theory.

The weak-binding limit is then defined as 
\begin{align}
   B
   &\ll E_{\rm typ} ,
   \label{eq:weakbindingE}
\end{align}
and we consider the expansion of compositeness in powers of $B/E_{\rm typ}$. For later convenience, we introduce the corresponding scales in the dimension of length as
\begin{align}
   R
   &=\frac{1}{\sqrt{2\mu B}} ,\quad
   R_{\rm typ}
   =\frac{1}{\sqrt{2\mu E_{\rm typ}}} .
   \label{eq:radius}
\end{align}
Then the weak-binding limit is also understood as
\begin{align}
   R
   &\gg R_{\rm typ} .
   \label{eq:weakbindingR}
\end{align}
Because $B\propto R^{-2}$, we have the relation
\begin{align}
   \sqrt{\frac{B}{E_{\rm typ}}}=\frac{R_{\rm typ}}{R} .
\end{align}

\subsection{Weak-binding limit}\label{subsec:wbl}

We now consider the weak-binding limit of the expression of the compositeness in Eq.~\eqref{eq:Xatpole}. When Eq.~\eqref{eq:weakbindingE} holds, the energy integration in Eq.~\eqref{eq:Xatpole} should be dominated by the $E\lesssim E_{\rm typ}$ region, because the integrand has a pole at $E=-B$. We then expand the form factor squared in powers of $E/E_{\rm typ}$:
\begin{align}
   |F(E)|^{2}
   &=
   1+c \frac{E}{E_{\rm typ}}+\mathcal{O}\left(\left(\frac{E}{E_{\rm typ}}\right)^{2}\right) ,\nonumber \\
   c&=E_{\rm typ}\left.\frac{d|F(E)|^{2}}{dE}\right|_{E=0} ,
   \label{eq:F2expansion}
\end{align}
where we have used the normalization $F(0)=1$. The expansion coefficient $c$ is dimensionless, and without any fine-tuning of the parameters, we expect $c\sim \mathcal{O}(1)$. Then the compositeness is evaluated as
\begin{align}
   X
   &\approx
   4\pi\sqrt{2\mu^{3}}g_{\rm th}^{2}\int_{0}^{E_{\rm typ}} dE
   \frac{\sqrt{E}[1+c E/E_{\rm typ}+\dotsb ]}{(E+B)^{2}} 
   \nonumber \\
   &= 4\pi\sqrt{2\mu^{3}}g_{\rm th}^{2}
   \frac{1}{\sqrt{B}}
   \left[
   \frac{\pi}{2}+\mathcal{O}\left(
   \left(\frac{B}{E_{\rm typ}}\right)^{1/2}
   \right)\right] .
\end{align}
We thus obtain the weak-binding limit expression of the compositeness
\begin{align}
   X
   &
   = 2\pi^{2}\sqrt{2\mu^{3}}
   \frac{g_{\rm th}^{2}}{\sqrt{B}}
   \left[
   1+\mathcal{O}\left(
   \left(\frac{B}{E_{\rm typ}}\right)^{1/2}
   \right)\right] 
   \label{eq:XatpoleWB} .
\end{align}
This expression indicates that, in the $B\to 0$ limit, the coupling constant should behave as $g_{\rm th}^{2}\sim \sqrt{B}\mu^{-3/2}$ and finally vanish in order to give finite $X$. In addition, the weak-binding limit does not depend on the explicit value of $E_{\rm typ}$. 

It is instructive to evaluate the residue of the bound state pole~\eqref{eq:residue} in the weak-binding limit. Applying the expansion~\eqref{eq:F2expansion} for Eq.~\eqref{eq:residue}, we obtain
\begin{align}
   g^{2}
   &
   =g_{\rm th}^{2}
   \left[
   1+\mathcal{O}\left(
   \frac{B}{E_{\rm typ}}
   \right)\right] .
   \label{eq:couplingrelation}
\end{align}
Defining the function $G(E)$ as
\begin{align}
   G(E)
   &=
   \int d\bm{p}\frac{|F(E_{p})|^{2}}{E-E_{p}+i0^{+}} ,
   \label{eq:GfnQM}
\end{align}
from Eq.~\eqref{eq:Xatpole}, we can express the compositeness $X$ by the residue $g^{2}$ and the energy derivative of the function $G(E)$ at the bound state pole:
\begin{align}
   X
   &=
   -g^{2}\left.\frac{dG(E)}{dE}\right|_{E=-B}
   \left[
   1+\mathcal{O}\left(
   \frac{B}{E_{\rm typ}}
   \right)\right] .
   \label{eq:compositenessresidue}
\end{align}
This corresponds to the expression of the compositeness evaluated at the bound state pole~\cite{Hyodo:2011qc,Aceti:2012dd,Sekihara:2014kya}. Further discussion will be given in Sect.~\ref{subsec:expressionsEFT}. Equations~\eqref{eq:XatpoleWB} and \eqref{eq:couplingrelation} also indicate that 
\begin{align}
   X
   &
   = 2\pi^{2}\sqrt{2\mu^{3}}
   \frac{g^{2}}{\sqrt{B}}
   \left[
   1+\mathcal{O}\left(
   \left(\frac{B}{E_{\rm typ}}\right)^{1/2}
   \right)\right] ,
   \label{eq:XatpoleWB2}
\end{align}
where the compositeness $X$ of an $s$-wave bound state is expressed only by the model-independent quantities, the binding energy $B$ and the residue of the pole $g^{2}$ in the weak-binding limit. 

Let us examine the bound state in higher partial waves. Performing the same expansion for Eq.~\eqref{eq:Xatpoleell} as $\ell\geq 1$, we obtain
\begin{align}
   X
   &= 4\pi\sqrt{2\mu^{3}}g_{\rm th}^{2}
   (2\mu)^{\ell}
   \frac{(E_{\rm typ})^{\ell-1/2}}{\ell-1/2} \nonumber \\
   &\quad \times \left[1+\mathcal{O}\left(
   \left(\frac{B}{E_{\rm typ}}\right)^{1/2}
   \right)\right]
   \quad (\ell\geq 1) ,
\end{align}
which is crucially different from Eq.~\eqref{eq:XatpoleWB} for $\ell=0$. In contrast to the $s$-wave case, the coupling constant $g_{\rm th}^{2}$ does not need to vanish in the weak-binding limit $B\to 0$, and the compositeness depends on the value of $E_{\rm typ}$. In fact, for the original integral~\eqref{eq:Xatpoleell} to converge, the form factor $|F(E)|^{2}$ should decrease as $E^{-\ell}$ at large $E$, and the value of the compositeness $X$ depends on the large $E$ behavior of the form factor. As mentioned above, the form factor reflects the off-shell behavior of the wave function
and is a model-dependent quantity. In this way, the compositeness $X$ contains model-dependent quantities for the bound state with $\ell\geq 1$, while for $s$-wave bound states, $X$ can be expressed by the model-independent quantities as in Eq.~\eqref{eq:XatpoleWB2}.

\subsection{Low-energy behavior of scattering amplitude}

Next we derive the expression of the compositeness in terms of the scattering parameters by considering the low-energy behavior of Eq.~\eqref{eq:scatteringamplitude}. Under the weak-binding situation $B\ll E_{\rm typ}$, we consider the low-energy scattering with $E\sim B\ll E_{\rm typ}$. In this case, the expansion of the form factor in Eq.~\eqref{eq:F2expansion} leads to
\begin{align}
   |F(E)|^{2}
   &
   \sim 1+ \mathcal{O}\left(\frac{B}{E_{\rm typ}}\right)
   \quad (E\sim B) .
\end{align}
In addition, Eq.~\eqref{eq:XatpoleWB} indicates $g_{\rm th}^{2}\sim \sqrt{B}\mu^{-3/2}$, the bound state pole term in Eq.~\eqref{eq:scatteringamplitude} should behave as
\begin{align}
   \frac{g_{\rm th}^{2}|F(E)|^{2}}{E+B}
   &
   \sim \frac{1}{\mu^{3/2} \sqrt{B}} \quad (E\sim B) ,
   \label{eq:poleterm}
\end{align}
where $\mu$ dependence is introduced for the dimensional reason. Equation~\eqref{eq:poleterm} indicates that this term is enhanced in the weak-binding limit. In contrast, the matrix element $v=\bra{\bm{p}}V\ket{\bm{p}}$ in Eq.~\eqref{eq:scatteringamplitude} is considered to be independent of the binding energy $B$, and therefore we assume that
\begin{align}
   v
   &
   \sim \frac{1}{\mu^{3/2} \sqrt{ E_{\rm typ}}} .
\end{align}
Under this assumption, $v$ can be neglected with respect to the pole term $g_{\rm th}^{2}|F(E)|^{2}/(E+B)$. Because the last integral in Eq.~\eqref{eq:scatteringamplitude} gives the contribution on the unitarity cut, it cannot be neglected to guarantee the conservation of probability. Thus, the approximation of Eq.~\eqref{eq:scatteringamplitude} in the low-energy region is given by
\begin{align}
   t(E)
   &
   =
   \left[\frac{g_{\rm th}^{2}}{E+B}
   +4\pi\sqrt{2\mu^{3}}\int_{0}^{\infty}dE^{\prime}
   \frac{\sqrt{E^{\prime}}|t(E^{\prime})|^{2}}{E-E^{\prime}+i0^{+}}
   \right] \nonumber \\
   & \quad \times \left[1+\mathcal{O}\left(
   \left(\frac{B}{E_{\rm typ}}\right)^{1/2}
   \right)\right].
   \label{eq:tlowenergy}
\end{align}
By neglecting the $\mathcal{O}((B/E_{\rm typ})^{1/2})$ terms, this integral equation can be solved analytically~\cite{Weinberg:1965zz}, thanks to the neglect of the matrix element $v$. In the framework of the two-component Hamiltonian in Eq.~\eqref{eq:twocomponent}, this corresponds to neglect the interaction $V$ in the diagonal component. In this case, the two-body interaction is solely given by the bare state pole term which is separable, and the Lippmann-Schwinger equation can be solved analytically~\cite{Hyodo:2014bda}. In the framework of effective field theory, this is equivalent to the resonance model with the vanishing four-point vertex~\cite{Kinugawa:2022fzn}, where the scattering amplitude equivalent to Eq.~\eqref{eq:fpWB} below is obtained in the small momentum limit.

The solution of Eq.~\eqref{eq:tlowenergy} can be translated to the scattering amplitude $f(p)=-4\pi^{2}\mu t(E_{p})$ as
\begin{align}
   f(p)
   &
   =
   \Biggl[
   -\frac{B}{4\pi^{2}\mu g_{\rm th}^{2}}
   -\frac{\sqrt{2\mu B}}{2} \nonumber\\
   &\quad +\frac{1}{2}
   \left(-\frac{1}{4\pi^{2}\mu^{2}g_{\rm th}^{2}}
   +\frac{1}{\sqrt{2\mu B}}\right)p^{2}
   -ip
   \Biggr]^{-1} ,
   \label{eq:fpWB}
\end{align}
where the denominator of the scattering amplitude is given by the quadratic function of $p$. Comparing it with the effective range expansion of the scattering amplitude
\begin{align}
   f(p)
   &
   =
   \Biggl[
   -\frac{1}{a_{0}}
   +\frac{r_{e}}{2}p^{2}
   +\mathcal{O}(p^{4})
   -ip
   \Biggr]^{-1} ,
   \label{eq:fpERE}
\end{align}
we obtain the expressions of the scattering length $a_{0}$ and effective range $r_{e}$ by $g_{\rm th}$ and $B$. With Eq.~\eqref{eq:XatpoleWB}, $a_{0}$ and $r_{e}$ can be expressed by the compositeness $X$ as 
\begin{align}
   a_{0}
   &
   =
   \frac{2X}{X+1}R+\mathcal{O}(R_{\rm typ})
   =
   \frac{2(1-Z)}{2-Z}R+\mathcal{O}(R_{\rm typ}), 
   \label{eq:a0WBR}\\
   r_{e}
   &
   =
   \frac{X-1}{X}R+\mathcal{O}(R_{\rm typ})
   =
   \frac{-Z}{1-Z}R+\mathcal{O}(R_{\rm typ}) .
   \label{eq:reWBR}
\end{align}
These are the weak-binding relations which express the scattering length $a_{0}$ and effective range $r_{e}$ by the compositeness $X$ and $R$ in Eq.~\eqref{eq:radius}. The model dependence of the compositeness disappears in Eq.~\eqref{eq:tlowenergy} where $v$ is neglected with respect to $g_{\rm th}^{2}/(E+B)$, and hence the compositeness is related to the observables. 

In this derivation, the relation of $a_{0}$~\eqref{eq:a0WBR} and the relation of $r_{e}$~\eqref{eq:reWBR} are obtained simultaneously. As shown in Ref.~\cite{Kamiya:2016oao}, in the framework of effective field theory, Eq.~\eqref{eq:a0WBR} is more fundamental, and Eq.~\eqref{eq:reWBR} is obtained by assuming the validity of the effective range expansion at the bound state pole. In fact, if the magnitude of the effective range is larger than the interaction range, the typical scale of the system $R_{\rm typ}$ in Eq.~\eqref{eq:a0WBR} should be estimated by the magnitude of the effective range to include the finite-range correction~\cite{Kinugawa:2022fzn}. Details will be discussed in Sect.~\ref{sec:EFT}.

In this way, it is shown that the compositeness $X$ of the $s$-wave bound state can be determined in a model-independent
manner. It should  however be kept in mind that we cannot completely exclude the possibility of fine-tuning to disturb this argument. For instance, if we artificially adjust the potential $V$ such that $v=\braket{\bm{p}|V|\bm{p}}\sim \mu^{-3/2}B^{-1/2}$ along with the $B\to 0$ limit, then Eq.~\eqref{eq:tlowenergy} does not hold. As is discussed in the ``Note added'' of the original paper~\cite{Weinberg:1965zz}, it is possible to have a CDD zero~\cite{Castillejo:1956ed,PR124.264,Baru:2010ww,Kamiya:2017pcq} in Eq.~\eqref{eq:tlowenergy}, and if the zero exists between the bound state pole and the threshold, then the above argument does not hold. In principle, these possibilities cannot be excluded. Nevertheless, it is expected that these cases are rarely realized. First of all, we recall that having a weakly bound state $B\ll E_{\rm typ}$ already requires one fine-tuning in the system. Namely, there exists one unnatural scale $R$ compared with the natural scale of the system $R_{\rm typ}$~\cite{vanKolck:2022lqz}. Thus, we need additional fine-tuning to realize $v\sim \mu^{-3/2}B^{-1/2}$ or to have a CDD zero at $E_{c}\ll E_{\rm typ}$. This double fine-tuning is expected to be highly unlikely (see also~Refs.~\cite{Hanhart:2014ssa,Sazdjian:2022kaf,Lebed:2022vks,Hanhart:2022qxq,Kinugawa:2023fbf}). In this sense, the weak-binding relations~\eqref{eq:a0WBR} and \eqref{eq:reWBR} are valid, provided that there is no further fine tuning on top of $B\ll E_{\rm typ}$.

\subsection{Discussion}

Before closing this section, let us make a few remarks on the compositeness. The compositeness $X$ can be defined for a bound state in any partial waves using Eq.~\eqref{eq:compositeness}, while it is a model-dependent quantity as discussed in Sect.~\ref{subsec:modeldependence}. What we show in Sect.~\ref{subsec:wbl} is that the compositeness can be determined in a model independent mannter for a weakly bound $s$-wave state. In other words, the model dependence of $X$ becomes small in the weak-binding limit.  This can be seen in Eqs.~\eqref{eq:a0WBR} and \eqref{eq:reWBR}; the model dependence of $X$ is compensated by the $\mathcal{O}(R_{\rm typ})$ terms whose magnitude is much smaller than the term proportional to $R$ in the weak-binding limit. This point will be further discussed in Sect.~\ref{sec:EFT}.

The peculiarity of the $s$-wave bound state is related to the low-energy universality~\cite{Braaten:2004rn,Naidon:2016dpf}. It is known that the scale-invariant zero-range limit is realized only in the $\ell=0$ case for three spatial dimensions~\cite{Nishida:2011ew}. This can be intuitively understood by the behavior of the partial wave scattering amplitude at small $p$:
\begin{align}
   f_{\ell}(p)
   &
   =\frac{p^{2\ell}}{-\frac{1}{a_{\ell}}+\frac{r_{\ell}}{2}p^{2}+\mathcal{O}(p^{4})-ip^{2\ell+1}} .
\end{align}
At small momentum $p\to 0$, the $s$-wave amplitude is characterized only by the scattering length $a_{0}$, while for $\ell\geq 1$ the  $p^{2}$ term cannot be neglected with respect to $-ip^{2\ell+1}$ so that there are at least two length scales $a_{\ell}$ and $r_{\ell}$ in the low-energy limit. The same observation can be made in the small $p$ behavior of the Jost function~\cite{Taylor,Hyodo:2014bda}. It is also known that the low-energy universality demands the relation for the $s$-wave bound state between the binding energy $B$ and the corresponding scattering length $a_{0}$ in the $B\to 0$ limit~\cite{Braaten:2004rn,Naidon:2016dpf}:
\begin{align}
   B
   &
   =\frac{1}{2\mu a_{0}^{2}} ,
\end{align}
which corresponds to $a_{0}=R$ and leads to $X=1$ in Eq.~\eqref{eq:a0WBR}. In fact, it is rigorously shown that $Z\to 0$ and $X\to 1$ in the strict $B\to 0$ limit, as long as the bound state pole remains in the scattering amplitude~\cite{Hyodo:2014bda}. In other words, the bound state becomes completely composite in the $B\to 0$ limit, irrespective of its origin. From this viewpoint, the weak-binding relation~\eqref{eq:a0WBR} represents the deviation from the strict $B\to 0$ limit by the coupling to other degrees of freedom ($X<1$ in the first term) and by the length scales in the system other than $a_{0}$ (the second term of order $\mathcal{O}(R_{\rm typ})$)~\cite{Kinugawa:2022fzn}.

In the coupled-channel scattering, if the bound state appears near the lowest-energy threshold, the weak-binding relation can be applied to the compositeness of the lowest energy channel $X_{1}$. Using the result of Refs.~\cite{Kamiya:2015aea,Kamiya:2016oao} to evaluate the contribution from the higher-energy channels, we can show that the scattering length of the lowest-energy channel $a_{0}$ can be given by
\begin{align}
   a_{0}
   &
   =
   R
   \left[\frac{2X_{1}}{X_{1}+1}
   +\mathcal{O}\left(\frac{R_{\rm typ}}{R}\right)
   +\mathcal{O}\left(\left(\frac{\ell}{R}\right)^{3}\right)
   \right] 
   \label{eq:a0WBRmulti} , \\
   R
   &=\frac{1}{\sqrt{2\mu_{1} B}} ,
   \quad 
   \ell
   =\frac{1}{\sqrt{2\mu_{1} E_{{\rm th},2}}} ,
\end{align}
where $\ell$ represents the length scale associated with the threshold energy of the second lowest energy channel. If the threshold energy of the channel 2 is much larger than the binding energy $B\ll E_{{\rm th},2}$, then we have
\begin{align}
   R \gg \ell,  
\end{align}
so that Eq.~\eqref{eq:a0WBRmulti} reduces to the single-channel version~\eqref{eq:a0WBR} for the compositeness of the lowest-energy channel $X_{1}$. This is consistent with naive expectation that the coupled-channel effect is suppressed if the threshold is far away from the energy region of interest. In this way, the compositeness of the lowest-energy channel $X_{1}$ can be determined independently of the model in the coupled-channel case. It should be emphasized that the compositeness $X_{i}$ for $i\geq 2$ cannot be model independently determined even if $B\ll E_{\rm typ}$, although $X_{i}$ can be defined as in Eq.~\eqref{eq:compositenessmulti}. If we define
\begin{align}
   X
   &
   =
   X_{1},
   \quad 
   Z 
   = \sum_{i=2}^{I}X_{i}+\sum_{n=1}^{N}Z_{n} ,
   \label{eq:coupledchannelZ}
\end{align}
then $X+Z=1$ follows thanks to Eq.~\eqref{eq:normalizationmulti}. While it is not possible to decompose $Z$ into various components, their sum in Eq.~\eqref{eq:coupledchannelZ} is model-independently determined in the weak-binding limit. This observation implies the generalized interpretation of the elementarity $Z$ in the weak-binding relation; $Z$ represents everything except for the compositeness of the lowest energy threshold. In this sense, ``elementarity'' is not a proper word to express $Z$, because $X_{2}, X_{3},\dotsb$ represent the fraction of composite components in higher energy scattering channels. 

In Table~\ref{tab:compositeness}, we summarize the definitions of the compositeness of a bound state  and their weak-binding limit in various cases. The compositeness can be defined as a model-dependent quantity in all cases for the state with an arbitrary binding energy. The compositeness of the lowest energy threshold becomes model-independent in the weak-binding limit, only for an $s$-wave bound state.

\begin{table}
\caption{Summary of definition of compositeness $X$ of a bound state and the weak-binding relation for various systems.}
\label{tab:compositeness}       
\begin{tabular}{lll}
\hline\noalign{\smallskip}
System & Definition & Weak-binding relation  \\
\noalign{\smallskip}\hline\noalign{\smallskip}
$\ell=0$, single channel     & $X$ in Eq.~\eqref{eq:compositeness} & Eq.~\eqref{eq:a0WBR} for $X$ \\
$\ell=0$, multi channel      & $X_{i}$ in Eq.~\eqref{eq:compositenessmulti} & Eq.~\eqref{eq:a0WBRmulti} for $X_{1}$ \\
$\ell\geq 1$, single channel & $X$ in Eq.~\eqref{eq:compositeness} & none \\
$\ell\geq 1$, multi channel  & $X_{i}$ in Eq.~\eqref{eq:compositenessmulti}  & none \\
\noalign{\smallskip}\hline
\end{tabular}
\end{table}

\section{Effective field theory}\label{sec:EFT}

In this section, we discuss the compositeness from the perspective of effective field theories (EFTs). Based on the principles established in Ref.~\cite{Weinberg:1979kz}, the EFT framework provides us with a systematic approach for describing low-energy phenomena of some microscopic theory. Thanks to the generality, EFTs have been applied in various fields of physics~\cite{Braaten:2007nq,Epelbaum:2008ga,Scherer:2012xha,Watanabe:2014fva,Hammer:2019poc}. 

Here we employ the non-relativistic EFT with contact interactions to describe the near-threshold dynamics of two-body systems. We follow the approach developed in Refs.~\cite{Kamiya:2015aea,Kamiya:2016oao,Kinugawa:2022fzn} where the ultraviolet cutoff $\Lambda$ is kept finite to represent the typical scale of the microscopic theory.\footnote{In this paper, we adopt the normalization~\eqref{eq:normalizationH0} also for the EFT. Because of this normalization, there are differences in the following expressions from those in Refs.~\cite{Kamiya:2015aea,Kamiya:2016oao,Kinugawa:2022fzn} by factors of $(2\pi)^{3}$ or $(2\pi)^{3/2}$. We summarize the conventions in~\ref{sec:convention}.} In other words, we consider that the microscopic theory has an intrinsic momentum scale $\Lambda$ as in Sect.~\ref{sec:WBR}, and we use the EFT to describe the low-momentum phenomena compared with $\Lambda$. Our aim is to derive the same weak-binding relation~\eqref{eq:a0WBR}, but with careful examination of the assumptions made.

Formally, the EFT used here are renormalizable, and the cutoff $\Lambda$ dependence can be absorbed by the bare coupling constants at each order under the proper power counting scheme~\cite{Hammer:2019poc}. The formulation of the compositeness in the renormalizable EFT is described in Ref.~\cite{vanKolck:2022lqz}. From this perspective, our formulation should be called ``EFT models.'' Nevertheless, we believe that the present formulation provides some insights on the results presented in Sects.~\ref{sec:formulation} and \ref{sec:WBR} from the field theory viewpoint.

\subsection{Hamiltonian and eigenstates}\label{subsec:HamiltonianEFT}

We consider a non-relativistic EFT with contact interactions. The minimal Hamiltonian to reproduce the setup in Sects.~\ref{subsec:Hamiltonian} is given by
\begin{align}
   H &=H_{0} + V 
   =  \int d\bm{r} (\mathcal{H}_{0}+\mathcal{V})\label{eq:Hamiltonian-bound}, \\
   \mathcal{H}_{0} 
   &=\frac{1}{2 M} \mathbf{\nabla} \psi^\dagger(\bm{r}) \cdot\mathbf{\nabla} \psi(\bm{r}) 
   +\frac{1}{2 m} \mathbf{\nabla} \phi^\dagger(\bm{r}) \cdot\mathbf{\nabla} \phi (\bm{r})
   \notag\\
   &\quad+ \frac{1}{2M_{0}} \mathbf{\nabla}  B_0^\dagger(\bm{r}) \cdot{\mathbf \nabla} B_0 (\bm{r})
   + \omega_0 B_0^\dagger(\bm{r}) B_0 (\bm{r}), \\
   \mathcal{V}
   &= g_{0}\left( B_0^\dagger(\bm{r}) \psi(\bm{r})\phi (\bm{r})
   + \phi^\dagger(\bm{r})\psi^\dagger (\bm{r})B_0(\bm{r}) \right) \nonumber \\
   &\quad +  v_{0} \psi^\dagger(\bm{r})\phi^\dagger (\bm{r})\phi(\bm{r})\psi(\bm{r})
   \label{eq:H_int} ,
\end{align}
where $\psi$ and $\phi$ are the fields for the two-body scattering channel and $B_{0}$ represents the bare field, $\omega_{0}$ stands for the bare energy of the $B_{0}$ field, and $g_{0}$ and $v_{0}$ are the bare coupling constants of the three-point and four-point vertices, respectively. The quantization condition for field $\alpha=\psi,\phi,B_{0}$ is given by 
\begin{align}
   [\alpha(\bm{r}),\alpha^{\dag}(\bm{r}^{\prime})]_{\pm}
   &= \delta(\bm{r}-\bm{r}^{\prime}) .
\end{align}

Let us construct the eigenstates of $H_{0}$. We first define the vacuum of the theory $\ket{0}$ as 
\begin{align}
   \tilde{\psi}(\bm{p})\ket{0} 
   = \tilde{\phi}(\bm{p})\ket{0} 
   = \tilde{B}_0(\bm{p})\ket{0}
   =0 ,\label{eq:vacuum}
\end{align}
with the normalization $\braket{0|0}=1$ and annihilation operators $\tilde{\alpha}(\bm{p})=\int d^3\bm{r} e^{-i\bm{p}\cdot\bm{r}}/(2\pi)^{3/2}\alpha(\bm{r})$. Then the eigenstates of $H_{0}$ are constructed as
\begin{align}
   \ket{\bm{p}}  
   = \frac{1}{\sqrt{ V_p}}\tilde{\psi}^\dagger(\bm{p}) \tilde{\phi}^\dagger(-\bm{p})\ket{0},
   \quad
   \ket{B_0}
   = \frac{1}{\sqrt{V_p}}\tilde{B}_0^\dagger(\bm{0})\ket{0},
   \label{eq:bareeigenstates}
\end{align}
with the phase space of the system $V_p = \delta^3(\bm{0})$. It can be shown that these states satisfy Eq.~\eqref{eq:SchroedingereqH0} with $\mu=Mm/(M+m)$~\cite{Kamiya:2016oao}. In this way, the bare state $\ket{B_{0}}$ can be explicitly constructed as an eigenstate of $H_{0}$ in this framework. In addition, the completeness relation~\eqref{eq:completenessH0} is also guaranteed by the number conservation which follows from the phase symmetry in the interaction~\eqref{eq:H_int}~\cite{Kamiya:2016oao}.

With these eigenstates, the matrix elements of $V$ can be obtained as
\begin{align}
   \braket{\bm{p}^{\prime}|V|\bm{p}}
   =\frac{v_{0}}{(2\pi)^{3}},
   \quad
   \braket{\bm{p}|V|B_{0}} = \frac{g_{0}}{(2\pi)^{3/2}} ,
   \label{eq:bareint}
\end{align}
which are constant in the momentum space, corresponding to the point-like interactions in the coordinate space. Such contact interactions are too singular at the origin, and regularization is needed for the problem to be well defined. We introduce an ultraviolet cutoff scale $\Lambda$ for the momentum integration, above which the interaction in the microscopic theory cannot be regarded as point-like. Namely, the cutoff $\Lambda$ is related to the typical length scale of the system $R_{\rm typ}$ in Eq.~\eqref{eq:radius} by
\begin{align}
   \Lambda \sim \frac{1}{R_{\rm typ}} .
   \label{eq:cutoff}
\end{align}
The regularization is achieved by replacing the matrix elements of $V$ as 
\begin{align}
   \braket{\bm{p}^{\prime}|V|\bm{p}}
   &=\frac{v_{0}}{(2\pi)^{3}}F(p^{\prime},\Lambda)F(p,\Lambda), \\
   \braket{\bm{p}|V|B_{0}} 
   &= \frac{g_{0}}{(2\pi)^{3/2}}F(p,\Lambda) ,
   \label{eq:vertexFF}
\end{align}
where $F(p,\Lambda)$ is a regulator~\cite{vanKolck:2022lqz} which satisfies
\begin{align}
   F(0,\Lambda)=1, \quad
   F(p,\Lambda)\to 0 \quad \text{for}\quad p\gg \Lambda .
\end{align}
For instance, a sharp cutoff
\begin{align}
   F(p,\Lambda)
   =\Theta(\Lambda-p) , \label{eq:sharpcut}
\end{align}
or the monopole (Yamaguchi) type~\cite{Yamaguchi:1954mp,Yamaguchi:1954zz} form factor
\begin{align}
   F(p,\Lambda)
   =\frac{\Lambda^{2}}{p^{2}+\Lambda^{2}} ,
   \label{eq:monopole}
\end{align}
are often used. Choosing a different regulator corresponds to the modification of the short range behavior of the microscopic interaction. Note that the introduction of the regulator $F(p,\Lambda)$ in Eq.~\eqref{eq:vertexFF} formally corresponds to dealing with the finite range interaction from the beginning. In this sense, our approach can be regarded as a generalization of the field theoretical models, such as the Lee model~\cite{Lee:1954iq,NC18.466,Vaughn:1961poz}, and the cutoff dependence and the regulator dependence represent the model dependence in this framework.

\subsection{Expressions of compositeness}\label{subsec:expressionsEFT}

When a bound state $\ket{B}$ exists as an eigenstate of the full Hamiltonian $H$, we can define the compositeness by Eq.~\eqref{eq:compositeness} in Sect.~\ref{subsec:definition}, but it is possible to express the compositeness in a different way in the EFT approach. Thanks to the number conservations, the two-body sector of this EFT is exactly solvable, and the on-shell $t$-matrix is obtained by solving the Lippmann-Schwinger equation as~\cite{Kamiya:2015aea,Kamiya:2016oao}
\begin{align}
   t(E)
   &
   =
   \frac{|F(p,\Lambda)|^{2}}{v^{-1}(E)-G(E)} ,
   \label{eq:tEFT}
\end{align}
with $p=\sqrt{2\mu E}$ and
\begin{align}
   v(E)
   &
   =
   \frac{1}{(2\pi)^{3}}
   \left[
   v_{0}+\frac{g_{0}^{2}}{E-\omega_{0}} 
   \right], 
   \label{eq:vfnEFT}\\
   G(E)
   &=
   \int d\bm{p}\frac{|F(p,\Lambda)|^{2}}{E-E_{p}+i0^{+}} .
   \label{eq:GfnEFT}
\end{align}
The function $v(E)$ corresponds to the effective interaction of the two-body system after eliminating the discrete channel $\ket{B_{0}}$ with the Feshbach method~\cite{Feshbach:1958nx,Feshbach:1962ut}. Namely, the first term $v_{0}$ represents the direct four-point interaction, and the second term corresponds to the interaction through the $s$-channel $B_{0}$ exchange, which is obtained as
\begin{align}
   \bra{\bm{p}^{\prime}}\left[
   V+\frac{V\ket{B_{0}}\bra{B_{0}}V}{E-\omega_{0}}
   \right]\ket{\bm{p}}
   &
   =v(E)F(p^{\prime},\Lambda)F(p,\Lambda)
\end{align}
The loop function $G(E)$ corresponds to the momentum representation of the free Green's function with the regulator $F(p,\Lambda)$, which corresponds to $G(E)$ in Eq.~\eqref{eq:GfnQM}. The same form of the scattering amplitude can be obtained by assuming the separable interaction model for $\braket{\bm{p}^{\prime}|V|\bm{p}}$~\cite{Sekihara:2014kya}. In the EFT approach, the contact interaction is separable from the beginning, which allows us to obtain the closed form solution of the Lippmann-Schwinger equation.

In this case, the coefficients $c$ and $\chi(\bm{p})$ of the wave function~\eqref{eq:twocomponent} can be determined by the Schr\"odinger equation and the normalization of the bound state wave function. The compositeness $X$ and the elementarity $Z$ are then expressed by $v(E)$ and $G(E)$ as~\cite{Kamiya:2015aea,Kamiya:2016oao}
\begin{align}
   X
   &
   =
   \left.
   \frac{G^{\prime}(E)}{G^{\prime}(E)-[1/v(E)]^{\prime}}
   \right|_{E=-B} , 
   \label{eq:compositenessGV}\\
   Z
   &=
   \left.
   \frac{-[1/v(E)]^{\prime}}{G^{\prime}(E)-[1/v(E)]^{\prime}}
   \right|_{E=-B} ,
   \label{eq:elementarityGV}
\end{align}
with 
\begin{align}
   A^{\prime}(E)
   &
   =
   \frac{dA(E)}{dE} .
\end{align}
This expression provides insight on the energy dependence of the interaction. Because $Z$ is proportional to the energy derivative of $1/v(E)$, we obtain
\begin{align}
   X
   &=1 ,\quad
   Z=0,
   \quad \text{for} \quad
   v^{\prime}(E)=0 .
\end{align}
Namely, an energy-independent interaction always gives a fully composite bound state~\cite{Sekihara:2016xnq,Oller:2017alp}. This is natural in the present setup, because in Eq.~\eqref{eq:vfnEFT}, the energy dependence of $v(E)$ arises from the coupling to the bare state $B_{0}$. To have an energy-independent interaction, the coupling $g_{0}$ must be switched off so that the composite state is obtained. This point is made clear on more general grounds in Refs.~\cite{Sekihara:2016xnq,Oller:2017alp}.

At this point, it is instructive to remark on the sign of the energy dependence of the interaction. We first note that the energy derivative of $G$ below the threshold is real and negative:
\begin{align}
   G^{\prime}(E)
   &=-\int d\bm{p}\frac{|F(p,\Lambda)|^{2}}{(E-E_{p})^{2}}
   < 0 \quad \text{for}\quad E<0,
   \label{eq:Gprime}
\end{align}
in analogy with the attractive second order perturbation of the ground state energy. From Eqs.~\eqref{eq:compositenessGV} and \eqref{eq:elementarityGV}, to obtain $X\geq 0$ and $Z\geq 0$, we must have 
\begin{align}
   [1/v(E)]^{\prime}
   &\geq 0 
   \quad
   \Leftrightarrow \quad
   v^{\prime}(E)
   \leq 0 
   \quad \text{at}\quad E=-B .
   \label{eq:energyderivative}
\end{align}
In other words, if we consider the interaction with $v^{\prime}(E)>0$ at $E=-B$, then we obtain $X>1$ and $Z<0$. The EFT interaction $v(E)$ in Eq.~\eqref{eq:vfnEFT} satisfies Eq.~\eqref{eq:energyderivative} as long as $g_{0}^{2}\geq 0$. The model with $g_{0}^{2}< 0$ is known as the resonance model with a ghost field~\cite{Kaplan:1996nv,Braaten:2007nq} where the bare state $B_{0}$ is introduced as a negative norm state. The negative elementarity $Z<0$ for the model with $g_{0}^{2}< 0$ (and hence $v^{\prime}(E)>0$ at $E=-B$) is then interpreted as the fraction of the negative norm state in the bound state wave function. 

The energy dependence of the potential is closely related to the normalization of the wave function of the bound state. It is known in quantum mechanics that to satisfy the continuity equation, the probability density $P(\bm{r})=|\psi_{E}(\bm{r})|^{2}$ should be modified for the energy~dependent potential $V(\bm{r},E)$ as~\cite{CJP54.289,Miyahara:2015bya,Miyahara:2018onh,Sekihara:2016xnq}
\begin{align}
   P(\bm{r})
   &=   \psi^{*}_{E}(\bm{r})
   \left[
   1-\frac{\partial V(\bm{r},E)}{\partial E}
   \right]
   \psi_{E}(\bm{r}) ,
\end{align}
where $\psi_{E}(\bm{r})$ is the wave function at energy $E$. Thus, the use of the energy-dependent potential leads to the modification of the  norm of the state. In the formulation of the compositeness in the EFT, this modification is interpreted as the elementarity $Z$, the fraction of the bare state component. The compositeness is defined as the norm of the two-body component $X=\int d\bm{r}|\psi_{E}(\bm{r})|^{2}$ at $E=-B$, with the normalization $\int d\bm{r}P(\bm{r})=1$. Then, one obtains the expression~\cite{Sekihara:2016xnq}
\begin{align}
   X
   &
   =\left.1+
   \int d\bm{r}\ \psi^{*}_{E}(\bm{r})
   \frac{\partial V(\bm{r},E)}{\partial E}
   \psi_{E}(\bm{r})
   \right|_{E=-B} .
\end{align}
Again, we should have $\partial V(\bm{r},E)/\partial E\leq 0$ for $X\leq 1$~\cite{Sekihara:2016xnq}. 

The compositeness $X$ in Eq.~\eqref{eq:compositenessGV} can also be expressed 
by the residue of the pole. Using Eq.~\eqref{eq:tEFT}, the residue of the bound state pole is expressed by $v$ and $G$ as
\begin{align}
   g^{2}
   &=
   \lim_{E\to -B}(E+B)t(E)
   =
   -\left.\frac{|F(p,\Lambda)|^{2}}{G^{\prime}(E)-[1/v(E)]^{\prime}}
   \right|_{E=-B}
   \label{eq:residueEFT} .
\end{align}
With the expression of the compositeness~\eqref{eq:compositenessGV}, we find
\begin{align}
   X
   &=-g^{2}G^{\prime}(E)|_{E=-B}
   \left[
   1+\mathcal{O}\left(
   \frac{B}{E_{\rm typ}}
   \right)\right]  ,
   \label{eq:compositenessresidueEFT}
\end{align}
which is equivalent to Eq.~\eqref{eq:compositenessresidue} in Sect.~\ref{subsec:wbl}. One notes that the terms of $\mathcal{O}(B/E_{\rm typ})$ stem from the regulator $F(p,\Lambda)$ reflecting the model dependence of the compositeness, which we discuss in the next section.


We show another expression of the compositeness $X$ by the self-energy $\Sigma(E)$, which appears in the following form of the on-shell $t$-matrix~\cite{Feshbach:1958nx,Feshbach:1962ut}:
\begin{align}
   t(E)
   &= \frac{|F(p,\Lambda)|^{2}}{(2\pi)^{3}v_{0}^{-1} - G(E)} \nonumber \\
   &\quad + \frac{{(2\pi)^{3}g_{0}^{2}v_{0}^{-2}[(2\pi)^{3}v_{0}^{-1}- G(E)}]^{-2}|F(p,\Lambda)|^{2}}{E - \omega_{0} - \Sigma(E)}, 
   \label{eq:tP+tV}\\
   \Sigma(E)
   &= \frac{g_{0}^{2}}{(2\pi)^{3}}\left[1 - \frac{v_{0}}{(2\pi)^{3}}G(E)\right]^{-1}G(E). 
\end{align}
The pole of $t(E)$ is given by the zero of the denominator of the second term of Eq.~\eqref{eq:tP+tV}, which is evolved from the bare state $B_{0}$ though the self-energy $\Sigma(E)$. With the energy derivative of $\Sigma(E)$, $X$ is written as~\cite{Hyodo:2014bda}
\begin{align}
   X
   &=\left.1-
   \frac{1}{1-\Sigma^{\prime}(E)}\right|_{E=-B} , \\
   Z
   &=\left.
   \frac{1}{1-\Sigma^{\prime}(E)}\right|_{E=-B}.
\end{align}
It can be shown that these are equivalent to Eqs.~\eqref{eq:compositenessGV} and \eqref{eq:elementarityGV}. 

\subsection{Loop function and model dependence}

In Sect.~\ref{subsec:modeldependence}, we emphasized the model dependence of the compositeness away from the weak-binding limit. Let us consider the same problem from the perspective of the EFT. For this purpose, the expression of $X$ by the residue $g^{2}$ and the derivative of the loop function~\eqref{eq:compositenessresidueEFT} is useful, because the residue $g^{2}$ is a model-independent quantity (see Table~\ref{tab:modeldep}). Thus, in this expression, the model dependence should be included in $G^{\prime}(E)$. As we already mentioned in Sect.~\ref{subsec:HamiltonianEFT}, the regulator $F(p,\Lambda)$ represents the short range behavior of the interaction in the microscopic theory, and a different choice of regulator modifies the value of the compositeness. 

To appreciate the model dependence, let us compute the loop function explicitly. With the sharp cutoff~\eqref{eq:sharpcut}, the loop function reads~\cite{Kinugawa:2023fbf}
\begin{align}
   G(E)
   &=
   -8\pi\mu
   \left[
   \Lambda + i\sqrt{2\mu E^{+}}\arctan\left(-\frac{\Lambda}{i\sqrt{2\mu E^{+}}}\right)\right] ,
\end{align}
where $E^{+}=E+i0^{+}$. Taking the energy derivative, we obtain
\begin{align}
   G^{\prime}(E)
   &=
   -8\pi\mu^{2}
   \Biggl[
   \frac{ i}{\sqrt{2\mu E^{+}}}
   \arctan\left(-\frac{\Lambda}{i\sqrt{2\mu E^{+}}}\right) \nonumber \\
   &\quad +
   \frac{\Lambda}{\sqrt{2\mu E^{+}}\left[1+\left(-\frac{\Lambda}{i\sqrt{2\mu E^{+}}}\right)\right]}
   \Biggr] .
\end{align}
With the monopole form factor~\eqref{eq:monopole}, the loop function and its derivative are given by
\begin{align}
   G(E)
   &=2\pi^{2}\mu\Lambda
   \frac{-\Lambda^{4}-2i\Lambda^{3}\sqrt{2\mu E^{+}}+2\Lambda^{2}\mu E}{(\Lambda^{2}+2\mu E)^{2}} ,\\
   G^{\prime}(E)
   &=
   -4\pi^{2}\mu^{2}\Lambda^{3}
   \frac{i\frac{\Lambda^{3}}{\sqrt{2\mu E^{+}}}-3\Lambda^{2}
   -3i\Lambda \sqrt{2\mu E^{+}}+2\mu E}{(\Lambda^{2}+2\mu E)^{3}} .
\end{align}
In this way, we see that the expression of $G^{\prime}(E)$, and hence the resulting $X$ in Eq.~\eqref{eq:compositenessresidueEFT}, depends on the choice of the regulator $F(p,\Lambda)$ and the value of the cutoff $\Lambda$. Physically speaking, this dependence is the remnant of the details of the short-range behavior of the microscopic interaction, which we call model dependence. 

For sufficiently low-energy $\sqrt{2\mu E^{+}}\ll \Lambda$, both expressions reduce to 
\begin{align}
   G(E)
   &=
   -C \Lambda
   -4\pi^{2}\mu i\sqrt{2\mu E^{+}}
   \left[1+\mathcal{O}\left(\frac{\sqrt{2\mu E^{+}}}{\Lambda}\right) \right] ,
\end{align}
with a constant $C$ which depends on the employed regulator.
This shows that the loop function is linearly divergent, and the imaginary part (which remains in the $\Lambda\to \infty$ limit) is completely determined by the unitarity. In this case, the derivative is given by
\begin{align}
   G^{\prime}(E)
   &=
   -i \frac{4\pi^{2}\mu^{2}}{\sqrt{2\mu E^{+}}}\left[1+\mathcal{O}\left(\frac{\sqrt{2\mu E^{+}}}{\Lambda}\right) \right] .
   \label{eq:Gprimewb}
\end{align}
Substituting this expression to Eq.~\eqref{eq:compositenessresidueEFT}, we arrive at Eq.~\eqref{eq:compositenessresidue}. Note that the leading order term is independent of $\Lambda$ and the explicit form of the regulator. This corresponds to the weak-binding situation where the model dependence is suppressed and the compositeness $X$ can be related to the observable quantities. 

From the perspective of the renormalization, one can let the bare parameters $v_{0},g_{0}$, and $\omega_{0}$ depend on the cutoff $\Lambda$ and consider the $\Lambda\to \infty$ limit. If there exits a sensible $\Lambda\to \infty$ limit with the physical observables being finite, the theory is called renormalizable~\cite{Braaten:2007nq}. This is the zero-range limit, where the interaction range is formally set to zero. In this case, one can neglect the higher-order terms in Eq.~\eqref{eq:Gprimewb} and obtain the compositeness
\begin{align}
   X
   &=g^{2} \frac{4\pi^{2}\mu^{2}}{\sqrt{2\mu B}} 
   \quad (\text{zero range limit}) ,
   \label{eq:Xzerorange}
\end{align}
which is also obtained by applying the dimensional regularization for $G(E)$. In this case, the compositeness is expressed only by the observable quantities, and model dependence disappears completely. We will come back to this point at the end of this section. 

\subsection{Weak-binding relations}

Here we consider the weak-binding relations~\eqref{eq:a0WBR} and \eqref{eq:reWBR} in the framework of the EFT. From the $t$-matrix in Eq.~\eqref{eq:tEFT} with $f(p)=-4\pi^{2}\mu t(E_{p})$, one can evaluate the scattering length $a_{0}$ by the threshold value of the scattering amplitude:
\begin{align}
   a_{0}
   &
   =
   4\pi^{2}\mu\frac{1}{1/v(0)-G(0)} .
\end{align}
With the identification of the typical scale by the cutoff as in Eq.~\eqref{eq:cutoff}, one can expand $a_{0}$ by $R_{\rm typ}/R$ to obtain~\cite{Kamiya:2015aea,Kamiya:2016oao}
\begin{align}
   a_{0}
   &
   =
   R
   \left[\frac{2X}{X+1}
   +\mathcal{O}\left(\frac{R_{\rm typ}}{R}\right)
   \right] 
   \label{eq:a0WBEFT} , 
\end{align}
which is equivalent to Eq.~\eqref{eq:a0WBR}. In this way, the weak-binding relation for $a_{0}$ is straightforwardly obtained in the EFT framework. 

By assuming that the bound state pole exists within the convergence radius of the ERE, we obtain relation between $a_{0}$, $r_{e}$ and $R$ as
\begin{align}
   -\frac{R}{a_{0}}-\frac{r_{e}}{2R}+1+\mathcal{O}\left(\left(\frac{R_{\rm typ}}{R}\right)^{3}\right) =0 .
   \label{eq:EREconvergence}
\end{align}
This bound state condition, together with the relation~\eqref{eq:a0WBEFT}, leads to the weak-binding relation for $r_{e}$:
\begin{align}
   r_{e}
   &
   =
   R
   \left[\frac{X-1}{X}
   +\mathcal{O}\left(\frac{R_{\rm typ}}{R}\right)
   \right] 
   \label{eq:reWBEFT} , 
\end{align}
which reproduces Eq.~\eqref{eq:reWBR}. Note that an additional assumption of the convergence of ERE at the bound state pole~\eqref{eq:EREconvergence} is needed to obtain the weak-binding relation~\eqref{eq:reWBEFT}, as mentioned earlier. 

It is shown~\cite{Kamiya:2016oao} that Eq.~\eqref{eq:reWBEFT} can also be obtained from Eq.~\eqref{eq:compositenessresidueEFT}. First, from Eqs.~\eqref{eq:residueEFT} 
we can express the residue $g^{2}$ by $R$ and $r_{e}$ as
\begin{align}
   g^{2}
   &=
   -\frac{1}{4\pi^{2}\mu}
   \lim_{p\to i/R}\frac{\left(\frac{p^{2}}{2\mu}+\frac{1}{2\mu R^{2}}\right)|F(p,\Lambda)|^{2}}{-\frac{1}{a_{0}}+\frac{r_{e}}{2}p^{2}-ip+\mathcal{O}(p^{4})} 
   \label{eq:g2ERE}\\
   &= \frac{1}{4\pi^{2}\mu^{2}}\frac{1}{R-r_{e}}\left[1
   +\mathcal{O}\left(\left(\frac{R_{\rm typ}}{R}\right)^{2}\right)
   \right] 
\end{align}
where we assume that the higher-order coefficients in the ERE has the natural size $\sim R_{\rm typ}$ which corresponds to the assumption of the convergence of the ERE. Combined with Eq.~\eqref{eq:Gprimewb}, we obtain the weak-binding relation~\eqref{eq:reWBEFT}. 

Let us now consider the zero-range limit where the correction terms $\mathcal{O}(R_{\rm typ}/R)$ can be neglected. In this case, the compositeness is expressed entirely by the observable quantities~\cite{Hyodo:2013iga,Oller:2017alp,Guo:2017jvc,Matuschek:2020gqe,Kinugawa:2022fzn}
\begin{align}
   X&=\frac{a_{0}}{2R-a_{0}} = \frac{1}{1-r_{e}/R} = \sqrt{\frac{1}{1-2r_{e}/a_{0}}} .
   \label{eq:XWBRzerorange}
\end{align}
These three expressions are equivalent if Eq.~\eqref{eq:EREconvergence} holds without the correction terms. In practice, we observe some deviation which stems from the correction terms. In addition, the expression by $r_{e}$ and $R$ shows that the compositeness exceeds unity ($X>1$) for positive effective range $r_{e}>0$. This contradicts the definition of real and nonnegative compositeness ($0\leq X\leq 1$). It is known that the effective range should be negative in the zero-range limit due to the Wigner bound~\cite{Phillips:1997xu,Oller:2017alp,Matuschek:2020gqe}. Thus, in the strict zero-range limit, $r_{e}$ must be negative such that $Z\leq 1$. In practice, the effective range of the $NN$ scattering in the $^{3}S_{1}$ channel is positive, $r_{e}\sim +1.8$ fm, and therefore naive application of the zero-range formula~\eqref{eq:XWBRzerorange} leads to the compositeness of the deuteron $X>1$. We will discuss this issue again in Sect.~\ref{subsec:deuteron}.

\subsection{Summary}

In this section, we have discussed the weak-binding relation for the compositeness in the framework of the EFT. The obtained relations~\eqref{eq:a0WBEFT} and \eqref{eq:reWBEFT} remain unchanged from Eqs.~\eqref{eq:a0WBR} and \eqref{eq:reWBR}, but the following points are better understood in the EFT framework:
\begin{itemize}

\item the eigenstates of the free Hamiltonian $H_{0}$ can be explicitly constructed in Eq.~\eqref{eq:bareeigenstates},

\item the typical length scale of the short-range interaction is naturally introduced by the ultraviolet cutoff~\eqref{eq:cutoff},

\item the elementarity is related to the energy dependence of the interaction in Eq.~\eqref{eq:elementarityGV}, and 

\item the assumptions needed to derive the weak-binding relations~\eqref{eq:a0WBEFT} and \eqref{eq:reWBEFT} can be systematically implemented.

\end{itemize}

\section{Unstable resonances}\label{sec:unstable}

So far we have concentrated on the compositeness of stable bound states. As mentioned in Sect.~\ref{sec:history}, however, for application to exotic hadrons, it is necessary to consider the compositeness of unstable resonances, because most hadrons are unstable against the strong decay. In this section, we first show that the natural generalization of the definition~\eqref{eq:compositeness} leads to the complex compositeness $X$. We then present several schemes proposed to interpret complex $X$. 

\subsection{Resonances and Gamow vectors}\label{subsec:gamow}

The concept of unstable states in the context of quantum mechanics was first described by Gamow in his theory of $\alpha$ decay of atomic nuclei~\cite{Gamow:1928zz}. Since then, mathematical formulation has been developed, and the unstable resonances can now be regarded as generalized eigenstates of the Hamiltonian with complex eigenenergy~\cite{Siegert:1939zz,Berggren:1968zz,Bohm:1981pv,Kukulin,Bohm:2001,delaMadrid:2002cz,Moiseyev,Hyodo:2020czb}. Let us first summarize the similarity and difference between stable bound states and  unstable resonances. The Schr\"odinger equation for a bound state $\ket{B}$ with the energy $E=-B$ is 
\begin{align}
   H\ket{B}
   &
   =
   -B\ket{B} ,
   \label{eq:boundSch}
\end{align}
where the binding energy $B$ is real. In the coordinate space wave function, we eliminate the exponentially growing component at large distance, which corresponds to imposing the purely outgoing boundary condition. By imposing the same outgoing boundary condition in the Schr\"odinger equation, an unstable resonance $\ket{R}$ is obtained as~\cite{Berggren:1968zz,Bohm:1981pv,Kukulin,delaMadrid:2002cz,Moiseyev,Hyodo:2020czb}
\begin{align}
   H\ket{R}
   &
   =
   E_{R}\ket{R}
   =
   \left(M_{R}-i\frac{\Gamma_{R}}{2}\right)\ket{R}, 
   \label{eq:resonanceSch}
\end{align}
where $M_{R}>0$ is the excitation energy and $\Gamma_{R}>0$ is the decay width. In this way, both $\ket{B}$ and $\ket{R}$ are (generalized) eigenstates of the Hamiltonian $H$, but the complex eigenenergy $E_{R}$ is associated with the resonance. It is shown that the eigenenergy in Eqs.~\eqref{eq:boundSch} and \eqref{eq:resonanceSch} corresponds to the pole of the scattering amplitude in the complex energy plane~\cite{Kukulin,Moiseyev,Hyodo:2020czb}.

There is an important difference in the normalization of the wave function. The bound state wave function asymptotically behaves as
\begin{align}
   \psi_{B}(\bm{r})=\braket{\bm{r}|B}
   \sim \frac{e^{-\sqrt{2\mu B}r}}{r} \quad (r\to \infty) .
\end{align}
Because it determines the dumping of the wave function, $R=1/\sqrt{2\mu B}$ is sometimes called the radius of the bound state. Thanks to this property, the wave function $\psi_{B}(\bm{r})$ is square integrable:
\begin{align}
   \braket{B|B}=\int d\bm{r}|\psi_{B}(\bm{r})|^{2}<\infty ,
\end{align}
and the normalization condition~\eqref{eq:normalizationfull} can be imposed. 

On the other hand, the asymptotic behavior of the resonance wave function $\psi_{R}(\bm{r})=\braket{\bm{r}|R}$ is
\begin{align}
   \psi_{R}(\bm{r})
   &\sim \frac{e^{i\sqrt{2\mu E_{R}}r}}{r}
   =\frac{e^{ip_{r}r}e^{+p_{i}r}}{r} \quad (r\to \infty), \\
   \sqrt{2\mu E_{R}}
   &=p_{R}-ip_{i} ,
\end{align}
with $p_{r}>0$ and $p_{i}>0$. Thus, the factor $e^{+p_{i}r}$ gives exponential growth at large distance, and the wave function is not square integrable
\begin{align}
   \braket{R|R}=\int d\bm{r}|\psi_{R}(\bm{r})|^{2}\to \infty .
\end{align}
In this case, the compositeness cannot be defined as in Sect.~\ref{subsec:definition}. 

In fact, the standard inner product $\braket{R|R}$ is not appropriate for normalizing the unstable resonance. For the resonance state satisfying Eq.~\eqref{eq:resonanceSch}, state $\bra{R}$ (conjugate of $\ket{R}$) is not an eigenstate of $H$ but $H^{\dag}$, with the different eigenenergy 
\begin{align}
   \bra{R}H^{\dag}
   &
   =
   \bra{R}E_{R}^{*}
   =
   \bra{R}\left(M_{R}+i\frac{\Gamma_{R}}{2}\right) .
\end{align}
Instead, we need to introduce the Gamow vector $\bra{\tilde{R}}$, which gives the same eigenvalue as $\ket{R}$:~\cite{Berggren:1968zz,Bohm:1981pv,delaMadrid:2002cz,Kukulin,Moiseyev,Hyodo:2013nka,Aceti:2014ala}
\begin{align}
   \bra{\tilde{R}}H
   &
   =
   \bra{\tilde{R}}E_{R}
   =
   \bra{\tilde{R}}\left(M_{R}-i\frac{\Gamma_{R}}{2}\right) .
   \label{eq:resonanceSchGamow}
\end{align}
In this case, it can be shown\footnote{The coordinate space representation of $\braket{\tilde{R}|R}$ still diverges in the usual sense, but it can be regularized by the Zel'dovich method~\cite{Berggren:1968zz}.} that $|\braket{\tilde{R}|R}|<\infty$ and a sensible normalization of the resonance state can be obtained by generalizing the inner product with $\bra{\tilde{R}}$ as~\cite{Berggren:1968zz}
\begin{align}
   \braket{\tilde{R}|R}
   &
   =1 .
   \label{eq:resonancenorm}
\end{align}
Mathematically, this is achieved with the biorthogonal basis formulated in the rigged Hilbert space~\cite{Bohm:1981pv,Bohm:2001}.

\subsection{Expressions of complex compositeness}\label{subsec:complex}

The generalized inner product with the Gamow vector gives the expectation value of the Hamiltonian
\begin{align}
   \bra{\tilde{R}}H\ket{R}
   &
   =E_{R}\in \mathbb{C} ,
\end{align}
which is consistent with Eqs.~\eqref{eq:resonanceSch}, \eqref{eq:resonanceSchGamow}, and \eqref{eq:resonancenorm}. This is a desired result because the complex eigenenergy is obtained by the expectation value. However, there is a price to pay: the expectation values of an operator, which is real for stable bound states, are in general complex. For instance, the mean squared radius of the resonance state becomes complex~\cite{Miyahara:2015bya}
\begin{align}
   \langle r^{2}\rangle
   &
   = \bra{\tilde{R}}r^{2}\ket{R}\in \mathbb{C}  .
\end{align}
Also, the compositeness $X$ (elementarity $Z$), which can be viewed as the expectation value of the  projection operator to the scattering states $P$ (to the bare state $Q$) defined in Eq.~\eqref{eq:projections}, is also complex:
\begin{align}
   X
   &
   =\bra{\tilde{R}}P\ket{R} 
   =\int d\bm{p}\ [\chi(\bm{p})]^{2} \in \mathbb{C} , 
   \label{eq:complexcompositeness}\\
   Z
   &= \bra{\tilde{R}}Q\ket{R} 
   =\braket{B_{0}|R}^{2} \in \mathbb{C} ,
\end{align}
because $\braket{\bm{p}|R}=\braket{\tilde{R}|\bm{p}}=\chi(\bm{p})$ and $\braket{B_{0}|R}=\braket{\tilde{R}|B_{0}}$~\cite{Berggren:1968zz}. We note that the normalization condition 
\begin{align}
   X
   +   Z
   &=1 ,
   \label{eq:normalization}
\end{align}
still holds for complex $X$ and $Z$. 

As in the same procedure to obtain Eq.~\eqref{eq:compositenessresidueEFT} for the bound state, the definition~\eqref{eq:complexcompositeness} can be rewritten by the derivative of the loop function in the second Riemann sheet $G_{\rm II}$ and the residue of the pole $g^{2}$ as~\cite{Hyodo:2011qc}
\begin{align}
   X
   &=-g^{2}G_{\rm II}^{\prime}(E)|_{E=E_{R}} \in \mathbb{C}.
   \label{eq:compositenessresidueresonance}
\end{align}
Note that $g^{2}$ is not the absolute value square but the complex number square, because the projection to the resonance state is given not by $\ket{R}\bra{R}$ but by $\ket{R}\bra{\tilde{R}}$~\cite{Kukulin}. 

For the coupled channel scattering, the compositeness of channel $i$ is defined as 
\begin{align}
   X_{i}
   &= \int d\bm{p}\ \braket{\tilde{R}|\bm{p},i}\braket{\bm{p},i|R} 
   =
   \int d\bm{p}\ [\chi_{i}(\bm{p})]^{2} ,
   \label{eq:compositenessmulticoupled} 
\end{align}
with $\braket{\bm{p},i|R}=\braket{\tilde{R}|\bm{p},i}=\chi_{i}(\bm{p})$. This can again be written by the residue of the pole as
\begin{align}
   X_{i}&=-g_{i}^{2}G_{i,\rm II}^{\prime}(E)|_{E=E_{R}} \in \mathbb{C}.
   \label{eq:compositenessresidueresonancecoupled}
\end{align}
For a near-threshold resonance in the coupled-channel scattering, the generalization of the weak-binding relation can be applied. Using the effective field theory, the generalized weak-binding relation is obtained as~\cite{Kamiya:2015aea,Kamiya:2016oao}
\begin{align}
   a_{0}
   &
   =
   R
   \left[\frac{2X}{X+1}
   +\mathcal{O}\left(
   \left|\frac{R_{\rm typ}}{R}\right|\right)
   +\mathcal{O}\left(
   \left|\frac{l}{R}\right|^{3}\right)
   \right] 
   \label{eq:a0WBEFTresonance} , 
\end{align}
where $l=1/\sqrt{2\mu \nu}$ is the length scale associated with the energy difference from the decay threshold $\nu$. Note that the scattering length $a_{0}$ and the length scale $R=\sqrt{2\mu E_{R}}$ are complex. In this case, the correction terms can be neglected if the magnitude of the eigenenergy $|E_{R}|$ is sufficiently small.

\subsection{Interpretation schemes}\label{subsec:interpretation}

Strictly speaking, the complex compositeness $X$ of unstable resonances cannot be interpreted as a probability~\cite{Hyodo:2011qc,Aceti:2014ala,Guo:2017jvc}. Yet, it is tempting to consider some prescription to gain insight into the structure of resonances by compositeness. In fact, there have been several prescriptions which can be classified into two approaches: either to calculate an interpretable quantity $X_{R}$ from complex $X$ in Eq.~\eqref{eq:complexcompositeness}, or to define a new quantity alternative to complex $X$. The latter approach was initiated in Refs.~\cite{Baru:2003qq,Kalashnikova:2009gt}, where the spectral density was used to define the real quantity corresponding to the elementarity $Z$. In Ref.~\cite{Tsuchida:2017gpb}, the real-valued compositeness is defined by considering the corresponding system in the finite volume. 

In the following, we focus on the former approach, and summarize the definitions of $X_{R}$ proposed so far. For simplicity, we consider the single channel case with Eq.~\eqref{eq:normalization}, but the generalization to the coupled channel case is straightforward. To establish a reasonable interpretation scheme, the quantity $X_{R}$ should satisfy several conditions based on the theoretical considerations. For instance, the following conditions are commonly introduced~\cite{Kamiya:2015aea,Kamiya:2016oao}:
\begin{itemize}
\item[i)] $0\leq X_{R}\leq 1$, and 
\item[ii)] $X_{R}$ reduces to $X$ for bound states.
\end{itemize}
Condition (i) guarantees that $X_{R}$ is regarded as a probability, and (ii) ensures the smooth bound state limit. As long as condition (ii) is respected, any interpretation schemes give a consistent result for narrow resonances. Condition (i) can be loosened if one considers that not all the states are interpretable. We emphasize here that this is a kind of interpretation problem; it is not possible to experimentally determine the superiority or inferiority of the various interpretation schemes.

A simple but natural prescription is to take the absolute value of the complex compositeness:
\begin{align}
   X_{R}=|X|
   \label{eq:CXabs}  .
\end{align}
In early works, this prescription is adopted heuristically~\cite{Aceti:2012dd,Hyodo:2013nka}. Later, a theoretically elaborated derivation of Eq.~\eqref{eq:CXabs} is given in Refs.~\cite{Guo:2015daa,Kang:2016ezb,Oller:2017alp} based on the convergence of the Laurent series expansion of the scattering amplitude. It is shown that a meaningful result can be obtained when the Laurent series around the resonance pole has an overlapping region with the real energy axis. This requires the additional condition ${\rm Re}\; E_{R}>0$, namely, the real part of the pole position should be above the threshold. Naturally, narrow resonances satisfy these conditions, but this scheme gives the meaningful relation even for broad resonances such as $f_{0}(500)=\sigma$~\cite{Guo:2015daa}, as long as proper conditions are satisfied.

In Ref.~\cite{Aceti:2014ala}, it is suggested to regard the real part of the complex compositeness as the ``weight'' of the composite component
\begin{align}
   X_{R}=\text{Re } X ,
   \label{eq:CXreal}
\end{align}
which corresponds to the integration of the real part of the wavefunction~\cite{Gamermann:2009uq,Aceti:2014ala}. In addition, Eq.~\eqref{eq:normalization} for complex $X$ and $Z$ indicates that the imaginary part of the left hand side vanishes as $\text{Im } X+\text{Im } Z=0$, and the sum rule for the real parts $\text{Re } X+\text{Re } Z = 1$ follows. 
We note that condition i) is not always satisfied by $X_{R}$ in Eqs.~\eqref{eq:CXreal} and \eqref{eq:CXabs}, but it is expected that condition i) is not violated for narrow resonances because of condition ii).

In fact, it is discussed in Ref.~\cite{Berggren:1970wto} that resonances with a broad decay width behave ambiguously, which is reflected in the imaginary part of the expectation value of an operator. This point is further discussed in relation to the compositeness in Refs.~\cite{Kamiya:2015aea,Kamiya:2016oao} by defining
\begin{align}
   X_{R}&= \tilde{X}=\frac{1-|Z|+|X|}{2}, 
   \label{eq:CXKH} \\
   U&= |Z|+|X|-1 .
   \label{eq:U}
\end{align}
Here, $X_{R}$ in Eq.~\eqref{eq:CXKH} always satisfies condition (i). The quantity $U$, motivated by $c_{n}$ in Ref.~\cite{Berggren:1970wto}, represents the uncertainty of the interpretation; $U=0$ in the limit of zero width, and $U$ increases when the imaginary part of $X$ is large. Thus, while Eq.~\eqref{eq:CXKH} gives interpretable $X_{R}$ for any states, those with highly unstable nature can be identified by large $U$. An alternative definition is proposed in Ref.~\cite{Sekihara:2015gvw} as
\begin{align}
   X_{R}&= \frac{|X|}{|X|+|Z|}
   = \frac{|X|}{1+U}, 
   \label{eq:CXSekihara}
\end{align}
which is advantageous for the generalization to coupled channel cases. It is explicitly demonstrated that the difference between all the above definitions of $X_{R}$ is equal to or smaller than $U/2$~\cite{Kamiya:2016oao}. This means that all the schemes agree with each other for the interpretation of narrow width resonances, which are expected to have similar properties to bound states. A direct application of the idea in Ref.~\cite{Berggren:1970wto} is presented in Ref.~\cite{Kinugawa:2024kwb},
where $X_{R}$ is given by
\begin{align}
X_{R}&=\mathcal{X}=\frac{(\alpha-1)|X|-\alpha|Z|+\alpha}{2\alpha-1}, \label{eq:calX} 
\end{align} 
with a real parameter $1/2<\alpha<\infty$. In addition, the elementarity $\mathcal{Z}$ and the uncertain probability $\mathcal{Y}$ are defined as 
\begin{align}
\mathcal{Y}&
=\frac{|X|+|Z|-1}{2\alpha-1}, \label{eq:calY} \\
\mathcal{Z}&
=\frac{(\alpha-1)|Z|-\alpha|X|+\alpha}{2\alpha-1}, \label{eq:calZ} 
\end{align} 
with $\mathcal{X}+\mathcal{Y}+\mathcal{Z}=1$. The probability $\mathcal{Y}$ corresponds to $U$ in Eq.~\eqref{eq:U} and $c_{n}$ in Ref.~\cite{Berggren:1970wto}, which is regarded as a probability in this scheme. In this scheme, $X_{R}=\mathcal{X}$ can be negative and condition i) is not always satisfied. It is suggested in Ref.~\cite{Kinugawa:2024kwb} to regard the states with $X_{R}<0$ as not interpretable. The parameter $\alpha$ determines the set of states with $X_{R}<0$. For instance, in the $\alpha\to 1/2$ limit, the interpretable states are restricted only to the stable bound states. The $\alpha\to \infty$ limit corresponds to Eq.~\eqref{eq:CXKH} where any states corresponding to a pole in the complex energy plane are interpretable. 

The interpretation of the compositeness in the zero-range limit~\eqref{eq:XWBRzerorange} is also discussed. It is proposed in Ref.~\cite{Hyodo:2013iga} to use the quantity 
\begin{align}
   \bar{X}&= \sqrt{\left|\frac{1}{1-2r_{e}/a_{0}}\right|} .
   \label{eq:CXH}
\end{align}
It is shown that the resonances in the $\Gamma_{R}\to 0$ limit are purely elementary with $X=0$, because it has no coupling to the scattering continuum. $\bar{X}$ in Eq.~\eqref{eq:CXH} measures the deviation from $X=0$ for finite-width resonances. In this case, condition i) is satisfied for resonances in the fourth quadrant in the second Riemann sheet of the complex energy plane, while for virtual states it gives $\bar{X}>1$. In Ref.~\cite{Matuschek:2020gqe}, based on the smoothness assumptions, the expression which is also applicable to the virtual state is proposed as
\begin{align}
   \bar{X}_{A}&= \sqrt{\frac{1}{1+|2r_{e}/a_{0}|}} .
   \label{eq:CXMatusschek}
\end{align}
Again, the quantities in Eqs.~\eqref{eq:CXH} and \eqref{eq:CXMatusschek} give consistent results with $X_{R}$ introduced above for narrow resonances~\cite{Kinugawa:2024kwb}.

\section{Applications}\label{sec:applications}

In this section, we summarize the results of the application of the compositeness to physical systems in the literature. In the following, we discuss the results of the selected states which are studied by many approaches. For applications to other states, see Refs.~\cite{Sekihara:2013wlq,MartinezTorres:2014kpc,Guo:2008zg,
Nagahiro:2011jn,
Xiao:2012vv,
Aceti:2014wka,
Sekihara:2015qqa,
Meissner:2015mza,
Navarra:2015iea,
Guo:2016wpy,
Lu:2016gev,
Kang:2016ezb}. 

\subsection{Compositeness of deuteron}\label{subsec:deuteron}

The deuteron is the prototype to which the compositeness is applied for the study of its internal structure. In the early studies in the 1960s, the internal structure of the deuteron is discussed by using the field renormalization constant $Z$. Note that the deuteron was considered a proton-neutron composite state even at that time, but it was used as a typical example for the theoretical approach to distinguish the elementary particle from the composite one, as mentioned in Sect.~\ref{sec:history}. In Ref.~\cite{PTP29.877}, for instance, the authors discuss the method to determine the composite or elementary from the experimental data, and they conclude the nonexistence of the elementary deuteron with great probability. Weinberg's work~\cite{Weinberg:1965zz} establishes the weak-binding relation and shows $Z\sim 0$ by the empirical values of the scattering length, effective range, and the binding energy.

The compositeness of deuteron was revisited recently in a more quantitative manner. It was however noted that the compositeness of the deuteron, estimated by the weak-binding relation in the zero-range limit~\eqref{eq:XWBRzerorange}, gives the result $X>1$. For instance, in Ref.~\cite{Kamiya:2017hni} the compositeness is evaluated as 
\begin{align}
   X=1.68^{+2.15}_{-0.83} ,
\end{align}
where the uncertainty is estimated by the correction terms $\mathcal{O}(R_{\rm typ}/R)$~\cite{Kamiya:2016oao}. As mentioned above, the compositeness in the zero-range limit~\eqref{eq:XWBRzerorange} exceeds unity for positive $r_{e}$, or equivalently, $a_{0}>R$. This is realized in the case of the deuteron, $a_{0}\sim 5.4$ fm, $r_{e}\sim 1.8$ fm, and $R\sim 4.3$ fm. This suggests that the finite-range correction is important to quantitatively analyze the deuteron. In studies with the range corrections, it is concluded that $\bar{X}_{A}=0.8$~\cite{Matuschek:2020gqe}, $X\geq 0.62$~\cite{Li:2021cue}, $X\sim 1$~\cite{Song:2022yvz}, $X\gtrsim 0.7 $ is very plausible~\cite{Albaladejo:2022sux}, $0.74\leq X\leq 1$~\cite{Kinugawa:2022fzn}, and $0.96\leq X\leq 1$~\cite{Oller:2022qau}. In this way, the composite dominance of the deuteron is now established quantitatively, including the finite range corrections. 

\subsection{Compositeness of $f_{0}(980)$ and $a_{0}(980)$}\label{subsec:scalar}

The lowest-lying scalar mesons, $f_{0}(500)=\sigma$, $\kappa$, $f_{0}(980)$ and $a_{0}(980)$, are considered to have an exotic internal structure due to their inverted spectrum from the naive $\bar{q}q$ assignment~\cite{Jaffe:2004ph}. Because $f_{0}(980)$ and $a_{0}(980)$ locate near the $K\bar{K}$ threshold, the structure of these scalar mesons has been studied by using the  $K\bar{K}$ compositeness. In the seminal work~\cite{Baru:2003qq}, Weinberg's method is generalized by using the spectral density, and it is concluded that the compositeness of $f_{0}(980)$ is 80 \% or more, and that of $a_{0}(980)$ is about 50-75 \%. In Refs.~\cite{Sekihara:2013wlq,Sekihara:2014kya,Sekihara:2014qxa}, the compositeness of $f_{0}(980)$ and $a_{0}(980)$ are evaluated from the residue of the pole~\eqref{eq:compositenessresidueresonancecoupled} in the chiral coupled-channel approaches. It is shown that $f_{0}(980)$ has appreciable $K\bar{K}$ compositeness, while $a_{0}(980)$ is not dominated by the $K\bar{K}$ component~\cite{Sekihara:2013wlq,Sekihara:2014kya}. The analysis of the $f_{0}(980)$-$a_{0}(980)$ mixing intensity indicates that $f_{0}(980)$ and $a_{0}(980)$ cannot be simultaneously $\bar{K}K$ molecular states, but the $f_{0}(980)$ has non-negligible $\bar{K}K$ component~\cite{Sekihara:2014qxa}.

References~\cite{Kamiya:2015aea,Kamiya:2016oao} estimate the $K\bar{K}$ compositeness of scalar mesons using the weak-binding relation~\eqref{eq:a0WBEFTresonance}. By using the results of various Flatt\'e-type analyses of experimental data, it is found that $a_{0}(980)$ is not dominated by the $K\bar{K}$ component, while $f_{0}(980)$ is composite dominant, although the results are not conclusive because the input data are somehow scattered. The evaluation of the compositeness of $f_{0}(980)$ from the residue of the pole~\eqref{eq:compositenessresidueresonancecoupled} is also performed in Ref.~\cite{Guo:2015daa}. Total compositeness is evaluated as $X=0.67^{+0.28}_{-0.27}$ which is almost saturated by the $K\bar{K}$ component ($X_{K\bar{K}}=0.65^{+0.27}_{-0.26}$). A detailed comparison of the compositeness from the residue of the pole~\eqref{eq:compositenessresidueresonancecoupled} and the integration of the spectral density is made in Ref.~\cite{Wang:2022vga}. In addition, the nature of the relation between the imaginary part of the pole position and the partial decay widths are clarified~\cite{Burkert:2022bqo}. It is found that $f_{0}(980)$ is dominated by the meson-meson components, in particular by the $\bar{K}K$ component, while the mesonic molecule components are subdominant in $a_{0}(980)$.

In summary, these studies consistently indicate that $f_{0}(980)$ is dominated by the $K\bar{K}$ molecular component, while the result of $a_{0}(980)$ depends on the analysis. This can be traced back to the relatively large uncertainty in the experimental data for $a_{0}(980)$. Precise data in future experiments is desired to pin down the nature of $a_{0}(980)$ by the compositeness.

\subsection{Compositeness of $\Lambda(1405)$}\label{subsec:L1405}

The $\Lambda(1405)$ resonance near the $\bar{K}N$ threshold is considered as a hadronic molecule state rather than the excited three-quark state~\cite{Hyodo:2011ur,Meissner:2020khl,Mai:2020ltx,Hyodo:2020czb,Sadasivan:2022srs}. Thanks to the developments of the next-to-leading order (NLO) chiral SU(3) dynamics ~\cite{Ikeda:2011pi,Ikeda:2012au,Guo:2012vv,Mai:2014xna}, the existence of two states, $\Lambda(1405)$ and $\Lambda(1380)$, has been established in this energy region~\cite{ParticleDataGroup:2024cfk}. This has been recently confirmed by the next-to-next-to-leading order (NNLO) analysis~\cite{Lu:2022hwm} as well as the meson-baryon scattering calculation on the lattice~\cite{BaryonScatteringBaSc:2023zvt,BaryonScatteringBaSc:2023ori}. 

The evaluation of the compositeness of $\Lambda(1405)$ and $\Lambda(1380)$ using the leading-order Weinberg-Tomozawa chiral models is performed in Ref.~\cite{Sekihara:2013wlq}. It is found that the $\bar{K}N$ component is dominant in $\Lambda(1405)$. The largest meson-baryon component in $\Lambda(1380)$ is $\pi\Sigma$, but the elementarity also occupies the substantial amount. This is consistent with the picture of $\Lambda(1405)$ as the $\bar{K}N$ quasi-bound state, and $\Lambda(1380)$ as the $\pi\Sigma$ resonance~\cite{Hyodo:2007jq}. The compositeness is also evaluated in Ref.~\cite{Garcia-Recio:2015jsa}, in the model with SU(6) extension, where the vector mesons and decuplet baryons are included in addition to the pseudoscalar mesons and octet baryons. The same tendency as in Ref.~\cite{Sekihara:2013wlq} is observed, presumably because the higher-energy meson-baryon channels in the SU(6) model do not contribute very much to the energy region of $\Lambda(1405)$ and $\Lambda(1380)$.

The compositeness of $\Lambda(1405)$ in the NLO chiral models~\cite{Ikeda:2011pi,Ikeda:2012au,Guo:2012vv} is evaluated from the residue of the pole~\cite{Sekihara:2014kya,Guo:2015daa}. Because the pole of $\Lambda(1405)$ is located near the $\bar{K}N$ threshold, the analysis with the weak-binding relation is also performed with the same scattering amplitudes~\cite{Kamiya:2015aea,Kamiya:2016oao}. It is instructive to compare the results with different procedures and with the different interpretation schemes. We summarize the $\bar{K}N$ compositeness in Table~\ref{tbl:compositeness}. The deviation of the result by the weak-binding relation from that with the residue of the pole is found to be smaller than 0.1, when the amplitude of Refs.~\cite{Ikeda:2011pi,Ikeda:2012au} is used. With the amplitude in Ref.~\cite{Guo:2012vv}, the weak-binding relation gives $|X|=0.92$, which is included in the uncertainty band given in Ref.~\cite{Guo:2015daa}. This indicates that the pole of $\Lambda(1405)$ is sufficiently close to the $\bar{K}N$ threshold, and the weak-binding relation gives the accurate estimation of the compositeness with the negligible correction terms. Namely, the model-dependence of the $\bar{K}N$ compositeness of $\Lambda(1405)$ is small due to the low-energy universality. The interpretation of the complex compositeness by $\tilde{X}$ in Eq.~\eqref{eq:CXKH} gives $\tilde{X}=1.0$ for the amplitude in Refs.~\cite{Ikeda:2011pi,Ikeda:2012au} and $0.9$ for that in Ref.~\cite{Guo:2012vv}. The difference between the scattering amplitude can be regarded as the systematic uncertainty in the experimental analysis, and the deviation is as small as 0.1. 

In this way, even with the model dependence and the interpretation schemes dependence, the $\bar{K}N$ compositeness of $\Lambda(1405)$ is close to unity, and it is concluded that $\Lambda(1405)$ is dominated by the  $\bar{K}N$ molecular component. Including the theoretical uncertainties in Table~\ref{tbl:compositeness}, the $\bar{K}N$ compositeness of $\Lambda(1405)$ is larger than 0.5, and these results establish the $\bar{K}N$ molecular picture for $\Lambda(1405)$. 


\begin{table}[bt]
\caption{$\bar{K}N$ compositeness of $\Lambda(1405)$ in the NLO chiral SU(3) amplitude from the residue of the pole~\eqref{eq:compositenessresidueresonancecoupled}~\cite{Sekihara:2014kya,Guo:2012vv} and by the weak-binding relation~\eqref{eq:compositenessresidueresonance}~\cite{Kamiya:2015aea,Kamiya:2016oao}.}
\begin{center}
\renewcommand{\arraystretch}{1.4}
\begin{tabular}{lll}
\hline
 Amplitude & Residue of the pole & Weak-binding relation \\ \hline
NLO~\cite{Ikeda:2011pi,Ikeda:2012au}  
  & $X=1.14+0.01i$
  & $X=1.2+0.1i$ \\
  & 
  & $\tilde{X}=1.0^{+0.0}_{-0.4}$ \\
  \hline
Fit II~\cite{Guo:2012vv} 
  & 
  & $X=0.9-0.2i$ \\
  & $|X|=0.82^{+0.36}_{-0.17}$ 
  & $\tilde{X}=0.9^{+0.1}_{-0.4}$ \\
\hline
\end{tabular}
\end{center}
\label{tbl:compositeness}
\end{table}%

\subsection{Compositeness of $X(3872)$}\label{subsec:X3872}

$X(3872)$ is a renowned exotic hadron candidate near the $D^{0}\bar{D}^{*0}$ threshold, firstly observed in the $J/\psi\pi^{+}\pi^{-}$ spectrum of the $B$ meson decay by the Belle collaboration~\cite{Belle:2003nnu,Hosaka:2016pey,Guo:2017jvc,Brambilla:2019esw,Chen:2022asf}. In early studies, the structure of $X(3872)$ is studied by the elementarity $Z$ in order to extract the $c\bar{c}$ core component. The elementarity is evaluated as $Z=0.31$-$0.37$ by the spectral density analysis of $B$ decay data by Belle~\cite{Kalashnikova:2009gt}. Inclusion of the $p_{T}$ distribution and total cross sections in the high energy collisions by CMS and CDF leads to $Z=0.28$-$0.44$~\cite{Meng:2013gga}. In Ref.~\cite{Chen:2013upa}, it is found from the analysis of the $D^{0}\bar{D}^{*0}$ spectrum in Belle and BABAR data that $Z=0.19\pm0.29$, indicating the nonvanishing $c\bar{c}$ core component. Using the Flatt\'e analysis of the LHCb data~\cite{LHCb:2020xds}, Ref.~\cite{Esposito:2021vhu} shows that $X(3872)$ contains non-negligible elementarity $0.052<Z<0.14$, also indicated by the large and negative effective range in Ref.~\cite{LHCb:2020xds} Note, however, that Ref.~\cite{Baru:2021ldu} pointed out the difficulty of determining the effective range in the experimental analysis.
In a detailed analysis in Ref.~\cite{Kang:2016jxw} using the amplitude with the CDD zero contributions, it is found that the compositeness is sensitive to the details of the pole structure around the threshold. 

In the recent analysis using the weak-binding relation~\eqref{eq:CXMatusschek}, the $D^{0}\bar{D}^{*0}$ compositeness of $X(3872)$ is estimated as $\bar{X}_{A}\gtrsim 0.9$~\cite{Baru:2021ldu}, and $0.53\leq X\leq 1$ including the range corrections~\cite{Kinugawa:2022fzn}. In the coupled-channel analysis, it is pointed out in Ref.~\cite{Kinugawa:2023fbf} that the $D^{0}\bar{D}^{*0}$ compositeness can be shared by the decay channel ($J/\psi\pi^{+}\pi^{-}$) and the coupled charged channel ($D^{*-}D^{+}$). By taking into account the LHCb data~\cite{LHCb:2020xds,Baru:2021ldu}, Ref.~\cite{Song:2023pdq} shows that the total compositeness (sum of neutral and charged channels) is of the order of 95 \%. In the meson exchange model coupled with the $c\bar{c}$ core~\cite{Wang:2023ovj}, the compositeness is evaluated as $X_{D^{0}\bar{D}^{*0}}=0.94$ and $X_{D^{*-}D^{+}}=0.048$, indicating the dominance of the molecular components. Using the coupled-channel EFT model in Ref.~\cite{Kinugawa:2023fbf} with the bare energy $\nu_{0}=78.36$ MeV from the constituent quark model~\cite{Godfrey:1985xj}, the complex compositeness of the neutral channel is found to be $X_{D^{0}\bar{D}^{*0}}=0.919-0.079i$. With the interpretation scheme using Eqs.~\eqref{eq:calX}, \eqref{eq:calY}, and \eqref{eq:calZ}, this gives $\mathcal{X}=0.890$, $\mathcal{Y}=0.028$, and $\mathcal{Z}=0.081$.

In summary, all these studies indicate that the structure of $X(3872)$ is dominated by the molecular component, mainly by the nearest $D^{0}\bar{D}^{*0}$ channel, with some fraction of the $D^{*-}D^{+}$ channel. This is expected from the mass of $X(3872)$ with respect to the threshold energies. At the same time, it is also shown that $X(3872)$ has nonnegligible elementarity $Z$, indicating the tiny fraction of the $c\bar{c}$ core component.

\subsection{Compositeness of $T_{cc}$}\label{subsec:Tcc}

The tetraquark $T_{cc}$ with charm $C=+2$ is observed in the high-energy $pp$ collisions near the $D^{0}D^{*+}$ threshold by the LHCb collaboration~\cite{LHCb:2021vvq,LHCb:2021auc}. In Ref.~\cite{Baru:2021ldu}, by applying weak-binding relation~\eqref{eq:CXMatusschek} to the threshold parameters in Ref.~\cite{LHCb:2021auc} with corrections, the molecular dominance of $T_{cc}$ is found with $\bar{X}_{A}\gtrsim 0.6$. Coupled-channel calculation including the three-body $D^{0}D^{0}\pi^{+}$ channel is performed in Ref.~\cite{Du:2021zzh}, and the central values of the compositeness are given as $\bar{X}_{A}= 0.84$-$0.88$ with the weak-binding relation~\eqref{eq:CXMatusschek}, and $X_{D^{0}D^{*+}}=0.71$-$0.73$ and $X_{D^{+}D^{*0}}=0.27$-$0.29$ from the residue of the pole~\eqref{eq:compositenessmulticoupled}. By using the weak-binding relation with the finite range correction, Ref.~\cite{Albaladejo:2022sux} shows that the structure of $T_{cc}$ is consistent with the molecular picture. In Ref.~\cite{Kinugawa:2023fbf} the compositeness of $T_{cc}$ is estimated under the variation of the EFT model parameters for the fixed binding energy. It is found that the probability of finding the composite dominant model is large, reinforcing the molecular dominance of $T_{cc}$. Based on the fit of the coupled-channel amplitude with contact interactions to the LHCb data, Ref.~\cite{Dai:2023cyo} reports the results $X_{D^{0}D^{*+}}=0.67$, $X_{D^{+}D^{*0}}=0.26$, and $Z=0.07$. The coupled-channel meson-exchange model, the compositeness is found to be $X_{D^{0}D^{*+}}=0.70$ and $X_{D^{+}D^{*0}}=0.30$~\cite{Wang:2023ovj}. 
With the quark model estimation of the bare energy $\nu_{0}=7$ MeV~\cite{Karliner:2017qjm}, we find the compositeness $X_{D^{0}D^{*+}}=0.541-0.007i$ in the coupled-channel model in Ref.~\cite{Kinugawa:2023fbf}. This result gives $\mathcal{X}=0.537$, $\mathcal{Y}=0.008$, and $\mathcal{Z}=0.456$. 

In this way, the studies of the compositeness of $T_{cc}$ indicate its molecular dominance. Quantitative difference of the evaluated compositeness can be traced back to the limited experimental data for $T_{cc}$, as pointed out in Ref.~\cite{Baru:2021ldu}. More experimental information on $T_{cc}$ is highly welcome to further pin down the nature of $T_{cc}$ quantitatively. 

\subsection{Other systems}\label{subsec:others}

Compositeness is also discussed in the systems other than hadrons. In nuclear physics, the structure of ${}^{3}_{\Lambda}$H is considered as a two-body system of deuteron and $\Lambda$. The $d\Lambda$ compositeness is estimated by the weak-binding relation as $0.74\leq X\leq 1$~\cite{Kinugawa:2022fzn}. In atomic physics, the ${}^{4}$He dimer is known as a shallow bound state of two ${}^{4}$He atoms. By using the weak-binding relation, the compositeness of ${}^{4}$He dimer is found to be $0.93\leq X\leq 1$, demonstrating the universal nature of the compositeness~\cite{Kinugawa:2022fzn}. In the cold atom physics, the wave-function renormalization factor $Z$ is used to characterize the molecular (closed-channel) fraction in the Bose gas~\cite{Duine:2003zza,Duine:2003zz,Braaten:2003sw,Naidon:2024bdy}, which corresponds to the elementarity $Z$ in Eq.~\eqref{eq:elementarity}.\footnote{Note that ``molecule'' in the atomic physics usually stands for the bare state component $\ket{B_{0}}$, not to the composite component which is called ``hadronic molecule'' in hadron physics. Hence, the ``molecular fraction'' corresponds to the elementarity $Z$.} Also, the compositeness and elementarity are used to characterize the quasi-particle weight of the molecule in the study of the polaron to molecule transition~\cite{Schmidt:2011zu}.

\section{Summary}\label{sec:summary}

In this paper, we have reviewed the formulation of the compositeness in detail. We show that the compositeness $X$ characterizes the fraction of the hadronic molecular component quantitatively. For a shallow bound state, the compositeness can be determined in a model-independent manner, thanks to the low-energy universality. Effective field theory framework is introduced to further provide insights in the formulation and the weak-binding relation of the compositeness. We survey the attempts to generalize of the compositeness to unstable resonances, which is needed for the application to exotic hadrons. As applications, the compositeness is found to be useful for studying the composite nature of the exotic hadrons, including light scalar mesons, $\Lambda(1405)$, $X(3872)$, and $T_{cc}$. Moreover, the universality allows us to consider the compositeness in different energy scales such as nuclei and atomic systems. 

\begin{acknowledgements}
The authors thank Shimpei Endo, Pascal Naidon, Jos\'e Antonio Oller and Eulogio Oset for fruitful discussions.
This work has been supported in part by the Grants-in-Aid for Scientific Research from JSPS (Grant numbers
No.~JP23KJ1796, 
No.~JP23H05439, 
JP22K03637, 
JP19H05150, 
JP18H05402). 
This work was supported by JST, the establishment of university
fellowships towards the creation of science technology innovation, Grant Number JPMJFS2139. 
\end{acknowledgements}

\appendix
\section{Conventions and normalization of states}\label{sec:convention}

In Sect.~\ref{sec:EFT}, we adopt the convention used in the non-relativistic quantum mechanics. The two-body scattering states are normalized as
\begin{align}
   \braket{\bm{p}|\bm{p}^{\prime}}
   &=\delta(\bm{p}-\bm{p}^{\prime}), \\
   1
   &= \int d\bm{p} \ket{\bm{p}}\bra{\bm{p}}, \\
   V_{p}
   &=\delta^3(\bm{0}) .
\end{align}
The Fourier transformation of the field $\alpha=\psi,\phi,B_{0}$ to the momentum representation $\tilde{\alpha}$ is given by
\begin{align}
   \alpha(\bm{r})
   &= \int d\bm{p} \frac{e^{i\bm{p}\cdot\bm{r}}}{(2\pi)^{3/2}}
   \tilde{\alpha}(\bm{p}) ,\\
   \tilde{\alpha}(\bm{p})
   &= \int d\bm{r}\frac{e^{-i\bm{p}\cdot\bm{r}}}{(2\pi)^{3/2}}
   \alpha(\bm{r}) ,
\end{align}
and the quantization of the fields is defined as
\begin{align}
   [\alpha(\bm{r}),\alpha^{\dag}(\bm{r}^{\prime})]_{\pm}
   &= \delta(\bm{r}-\bm{r}^{\prime}) ,
\end{align}
which leads to
\begin{align}
   [\tilde{\alpha}(\bm{p}),\tilde{\alpha}^{\dag}(\bm{p}^{\prime})]_{\pm}
   &= \delta(\bm{p}-\bm{p}^{\prime})  .
\end{align}
In this case, the scattering amplitude $f(p)$ is related to the T-matrix $t(E_{p})$ as~\cite{Taylor}
\begin{align}
   f(p)=-4\pi^{2}\mu t(E_{p}) .
\end{align}

With the convention used in Refs.~\cite{Kamiya:2015aea,Kamiya:2016oao,Kinugawa:2022fzn}, the corresponding expressions are given by
\begin{align}
   \braket{\bm{p}|\bm{p}^{\prime}}
   &=(2\pi)^{3}\delta(\bm{p}-\bm{p}^{\prime}), \\
   1
   &= \int \frac{d\bm{p}}{(2\pi)^{3}} \ket{\bm{p}}\bra{\bm{p}} \\
   V_{p}
   &=(2\pi)^{3}\delta^3(\bm{0}) \\
   \alpha(\bm{r})
   &= \int \frac{d\bm{p}}{(2\pi)^{3}} e^{i\bm{p}\cdot\bm{r}}
   \tilde{\alpha}(\bm{p}) ,\\
   \tilde{\alpha}(\bm{p})
   &= \int d\bm{r} e^{-i\bm{p}\cdot\bm{r}}
   \alpha(\bm{r}) \\
   [\alpha(\bm{r}),\alpha^{\dag}(\bm{r}^{\prime})]_{\pm}
   &= \delta(\bm{r}-\bm{r}^{\prime}) , \\
   [\tilde{\alpha}(\bm{p}),\tilde{\alpha}^{\dag}(\bm{p}^{\prime})]_{\pm}
   &= (2\pi)^{3}\delta(\bm{p}-\bm{p}^{\prime})  , \\
   f(p)
   &=-\frac{\mu}{2\pi} t(E_{p}) .
\end{align}
These are typically adopted in the EFT because it has smooth generalization to the convention in the relativistic quantum field theories. 


\end{document}